\documentclass{aa}  
\makeatletter
\def\linenumbers{\@ifnextchar[{\@linenumbers}{\@linenumbers[]}}
\def\@linenumbers[#1]{}
\def\nolinenumbers{}
\def\linenumber#1{}

\def\setrunninglinenumbers{}
\def\runninglinenumbers{}

\makeatother
\usepackage{graphicx}
\usepackage{natbib}
\usepackage{txfonts}
\usepackage{lipsum}
\usepackage{subcaption}        
\usepackage{tabularx}
\usepackage{amsmath}
\usepackage{graphicx}
\usepackage{booktabs}
\usepackage{multirow}
\usepackage{soul}

\usepackage{graphicx}
\usepackage{natbib}
\usepackage{color}
\usepackage{hyperref}
\usepackage{lscape}             
\usepackage{placeins} 
\hypersetup{
    colorlinks=true,
    linkcolor=blue,
    citecolor=blue,
    urlcolor=blue
}

\begin{document}

   \title{Velocity-Space Signatures of Energy Transfer for Ion-Acoustic Instabilities}


   \author{Mahmoud Saad Afify\inst{1, 2, 3}
        \and Kristopher G. Klein\inst{2}
        \and Mihailo M. Martinović\inst{2}
        \and Maria Elena Innocenti\inst{1}
        }

   \institute{Theoretische Physik I, Ruhr-Universität Bochum, Bochum, Germany
   \and Lunar and Planetary Laboratory, University of Arizona, Tucson, AZ 85721, USA
            \and Department of Physics, Faculty of Science, Benha University, Benha 13518, Egypt\\
            \email{Mahmoud.Ibrahim@ruhr-uni-bochum.de\\Mahmoud.Afify@fsc.bu.edu.eg}}


  \abstract
    {Observations by Parker Solar Probe (PSP) of electrostatic waves suggest that electrostatic instabilities, including the ion-ion-acoustic instability (IIAI) frequently observed in the inner heliosphere, play an important role in plasma heating and particle acceleration. }
   {Our aim is to explore the application of single spacecraft diagnostics to the IIAI, in anticipation of application to the current missions operating in the inner heliosphere, e.g. PSP and Solar Orbiter.} 
   {We apply the field-particle correlation (FPC) technique to fully kinetic simulations of IIAI. We characterize the conversion of energy between the electric field and particle species, allowing the differentiation between oscillatory and secular energy transfer to and from the particles and highlighting the role of resonant energy exchange.
We then identify the characteristic IIAI signatures for the proton and electron distributions, and relate them to our previous knowledge of IIAI onset and energy exchange mechanisms.}
   {Applying the FPC technique to our simulations run in parameters regime compatible with solar wind conditions, we have identified IIAI signatures that would enable efficient recognition of IIAI in observations. This task is left for future missions, since the time scale over which IIAI signatures develop is too fast for the sampling rates of current missions.}
   {}


   \maketitle

\section{Introduction}
\label{sec:intro}
Electrostatic waves are ubiquitously observed in  the solar wind and in planetary magnetospheres and ionospheres \citep{hollweg1975waves, stix1992waves, GaryBook, mangeney1999wind, Baumjohann2012}. 
These waves are signatures of instabilities often driven by energetic ion and electron beams, which can originate from several different processes, including magnetic reconnection \citep{Dai2021, Phan2022}, turbulence \citep{Marsch1991, Perrone2011, Valentini2011, Valentini2014}, and parametric decay instability \citep{gonzalez2023particle}.  
A typical example of such waves are ion-acoustic waves (IAWs), which have been observed in the solar wind for over four decades  \citep{Gurnett1977, Kurth1979, Gurnett1991, Pisa2021, Graham2021} and are commonly found near collisionless shocks \citep{Wilson2007, goodrich2019impulsively, Bolduu2024, Graham2025IAWs} and magnetic reconnection sites \citep{Liu2024a, Graham2025b, li2025role}. 

Interest in electrostatic modes in the solar wind has been renewed by Parker Solar Probe (PSP,~\citet{Fox2016}) observations suggesting that electrostatic modes may compete with electromagnetic ones especially at lower heliospheric distances, where the occurrence of electromagnetic modes such as whistler waves is reduced~\citep{Cattell2022, nair2025suprathermal}. 
\citet{Cattell2022} suggests that the reduced occurrence of whistler waves might result from competition with lower-threshold electrostatic instabilities.

At 55 $R_{\odot}$, PSP detected Doppler-shifted IAWs characterized by wide bandwidth and short duration,  wavevectors aligned with the background magnetic field and  linear polarization ~\citep{Mozer2020}.
The proton velocity distribution function (VDF) exhibited a core and a beam populations, suggesting that the ion-ion acoustic instability, IIAI~\citep{Gary1987, GaryBook}, is the most plausible driver of the IAWs in this regime. 
At 35 $R_{\odot}$, PSP observed a variety of IA modes around magnetic field switchback boundaries ~\citep{Mozer2020,tenerani2021evolution}. 
At nonlinear amplitudes, IA modes were found to result in electron and ion holes \citep{Mozer2021bb}. 

Long-lived, narrow-banded electrostatic packets persisting for hours and composed of coupled pairs of a few hundred Hz and a few Hz oscillations have been observed by PSP between 15 and 25 $R_{\odot}$.
These packets have been identified as IAWs \citep{mozer2021triggered, Mozer2023a}.
The phase velocities of these waves roughly matched the local ion-acoustic speed, and the proton distribution functions displayed a plateau at this velocity, which can be modeled as a core and beam combination of two relatively drifting bi-Maxwellians.
Furthermore, \cite{malaspina2024frequency}   observed frequent occurrences of frequency-dispersed ion acoustic waves at distances $r< 60 R_\odot$, and associated them to cold, impulsively accelerated proton beams near the ambient proton thermal speed.
The ubiquity observation of IAW motivates an investigation of what impact these instabilities can have on the thermodynamic state of the plasma.

IAWs satisfy the dispersion relation 
$\omega_{\mathrm{r}} \simeq k_{\|} c_{\mathrm{s}}$ \citep{stix1992waves}, 
where 
$c_{\mathrm{s}} = \sqrt{(3 k_{\mathrm{B}} T_{\|\mathrm{i}} + k_{\mathrm{B}} T_{\|\mathrm{e}})/m_{\mathrm{i}}}$ 
is the ion-acoustic speed and $m_{\mathrm{i}}$ is the ion mass and $k_\mathrm{B}$ is the Boltzmann constant. For electrostatic IAWs, resonant wave-particle interaction occurs through the Landau resonance condition 
$v_{\|} = v_{\mathrm{res}} = \omega/k$, 
where $v_{\|}$ denotes the particle parallel velocity with respect to the mean magnetic field (when present) and $v_{\mathrm{res}}$ is the resonance (phase) velocity of the wave. The direction of energy exchange is determined by the sign of the velocity-space gradient of the distribution function evaluated at the resonance speed. 
If $\partial f / \partial v_{\|} > 0$ at $v_{\mathrm{res}}$, particles lose energy to the wave, thereby driving the instability. 
Conversely, if $\partial f / \partial v_{\|} < 0$ at $v_{\mathrm{res}}$, particles gain energy from the wave, resulting in Landau damping. A system is unstable to the IIAI if the ion acoustic speed falls in the $\partial f / \partial v_{\|} > 0$ region of the proton beam population (a condition which depends on the relative values of the proton core/ beam drift speed and of the resonance velocity, which in turns depends critically on the electron temperature), in the absence of significant electron Landau damping~\citep{Gary1987, Gurnett1991}. The relation between proton and electron thermal and drift speeds and IIAI threshold is examined in details in~\citet{Afify2024, Afify2025}.   

In our previous work,~\citet{Afify2024}, we have investigated IIAI onset in parameter regimes comparable with the observations in~\citet{mozer2021triggered}. We have demonstrated that the solar wind parameters from~\citet{mozer2021triggered} are \textit{in the vicinity of the IIAI threshold}, but not in the unstable regime itself. In particular, we have succeeded in reproducing well the observed growth time (inverse of the growth rate) and electric field magnitude ($\tau \sim 10\;ms$ and $E\sim 19\;mV\;m^{-1}$ from our simulations after renormalizing) of the high frequency IIAI observed there, but only after (slightly) modifying key parameters such as the electron-to-core temperature ratio, the ratio between the parallel beam-to-core temperature, and the relative drift between the core and beam protons. 
In a follow up paper, \citet{Afify2025}, we have investigated the impact of non-thermal electron distributions on IIAI onset. We have shown that, while non-thermal electron distributions (and in particular the core-strahl distributions frequently observed in the solar wind) have an impact on the onset and the growth rate of the IIAI, it is not strong enough to push the plasma towards IIAI in the parameter ranges described in \citet{mozer2021triggered}.   

Characterizing energy transfer and dissipation mechanisms in collisionless plasmas is essential for understanding a wide range of astrophysical and heliospheric phenomena \citep{howes2024fundamental}. 
The field-particle correlation technique (FPC) \citep{Klein2016, howes2017diagnosing} provides a novel approach to directly measure collisionless energy transfer in plasmas from single-point measurements in space. 
This last aspect is of particular relevance for current solar wind missions like PSP and Solar Orbiter, which are exploring the young solar wind near the Sun \citep{muller2020solar}.
FPC computes the correlation between fluctuations in the electric field and the charged particle velocity distribution function (VDF), revealing at which velocities energy is gained or lost by a particle species. 
By isolating specific regions in velocity space where particles either gain or lose energy, the technique allows for the identification of specific damping and acceleration mechanisms.
This methodology has been applied to simulations of Landau damping  \citep{Klein2017a}, transit-time damping  \citep{huang2024velocity}, cyclotron damping \citep{klein2020diagnosing} and  shock-drift acceleration~\citep{howes2025velocity, juno2023phase} and to analysis of ion-scale instability in collisionless shock \citep{brown2023isolation}
In all the works mentioned above, the FPC techniques has been applied to simulations. However, the technique can be applied to spacecraft observations as well, and it has been used to identify electron Landau damping~\citep{Chen2019, afshari2021importance},  ion cyclotron damping from turbulence~\citep{Afshari2024}, and signatures of shock drift acceleration \citep{montag2025mms} in Magnetospheric MultiScale, MMS, observations. \citet{Verniero2021} uses downsampled simulation results to identify which spacecraft could be used to identify FPC  ion Landau damping signatures in solar wind~\citep{Verniero2021}. FPC has been applied to an analytical model for magnetic pumping \citep{montag2022field}. 
As the method can be applied to both simulations and observations, one can use signatures of specific mechanisms derived from simulations, where the plasma conditions and ensuing wave-particle interaction process are fully controlled, to identify specific dissipation and instability processes in spacecraft observations \citep{Verniero2021}.

The goal of this paper is to identify characteristic FPC signatures of the IIAI from fully kinetic simulations. 
The paper is organized as follows. We recap the linear instability analysis for the IIAI and describe our fully kinetic Vlasov simulations in Sec.~\ref{sec:linear-analysis}. The energy evolution in our simulations is examined in Sec.~\ref{ssec:method.simulation}, where we pay particular attention to the energy variation for the different particle species.  We apply the Field Particle Correlation technique to our simulations in Sec.~\ref{sec:FPC} and finally, we present a discussion and conclusions of this work in Section \ref{sec:discussion}.

\section{Linear Instability Analysis and Numerical Simulations}
\label{sec:linear-analysis}
We briefly recap the IIAI linear theory, which is used to validate our simulations. 
We consider a system comprising three species, core protons, beam protons, and background electrons.
All species have Maxwellian VDFs.
For collisionless unmagnetized plasma system the the 1D-1V Vlasov equation reads as:
\begin{equation}
\frac{\partial f_s}{\partial t}+v \frac{\partial f_s}{\partial x}-\frac{q_s}{m_s} \frac{\partial \varphi}{\partial x} \frac{\partial f_s}{\partial v}=0. \label{eq:vlasov}
\end{equation}
The electric potential $\varphi$ can be calculated using Poisson's equation
\begin{equation}
\frac{\partial^2 \varphi}{\partial x^2}=-\frac{1}{\epsilon_0} \sum_s q_s n_s. \label{eq:poisson}
\end{equation}
where the number density is given as $n_s=\int_{-\infty}^{\infty} f_s d v$, $f_s(v)$ is the VDF, with $s=c$ for core protons, $b$ for beam protons and $e$ for electrons; $m_s$ is the mass, and $q_s$ is the electric charge. 
The electrostatic potential and the electric field are related by $E=-\partial \varphi / \partial x$.
In this study, we initialize the core protons with a non-drifting Maxwellian, $V_{D,c}=0$, for all simulations, while both the beam protons and electrons have drifting Maxwellian distributions, expressed as:
\begin{equation}
f_{0,s}(v) = \frac{n_s}{\sqrt{2\pi} v_{th,s}} \exp\left(-\frac{(v - V_{D,s})^2}{2v_{th,s}^2}\right), \label{equation3}
\end{equation}
where $V_{D,s}$ and $v_{th,s} = \sqrt{T_s / m_s}$ represent the drift and thermal velocities, with temperatures $T_s$ given in energy units. 
The zero current condition demands $n_b V_{D,b}-n_e V_{D,e} =0$ when considering $V_{D,c}=0$ as the reference frame. 
The length, time, and velocity are normalized to the core Debye length $\lambda_{Dc} = \frac{v_{th,c}}{\omega_{pc}}$, the core plasma frequency $\omega_{pc}$, and the core thermal velocity $v_{th,c}$, respectively. 
We use a realistic mass ratio $m_i/m_e=1836$ and equal temperatures for the two proton populations $T_b/T_c=1$. 
Within the framework of standard linear Landau theory, the dispersion relation for this system is derived by considering a linear perturbation around a homogeneous equilibrium.  

Applying a standard plane-wave perturbation ansatz of the form $\sim e^{i(kx - \omega t)}$ to the linearized system yields the following electrostatic dispersion relation \citep{Gary1987, Afify2024}:
\begin{equation}
2 k^2 \lambda_{Dc}^2-\alpha_e Z\left(\zeta_e\right)-\alpha_c Z\left(\zeta_c\right)-\alpha_b Z\left(\zeta_b\right)=0, 
\label{DR}
\end{equation}
where $\alpha_j=\frac{n_j}{n_c} \frac{T_c}{T_j}$, $Z\left(\zeta_j\right)$ is the plasma dispersion function \citep{fried1961plasma}, $\zeta_j=\frac{\omega - k  V_{D,j}}{\sqrt{2} k v_{th,j}}$, $\omega=\omega_r+i \gamma$ is the complex frequency, and $k$ is the wavenumber.

\begin{figure}[h!]
\centering
\begin{minipage}[b]{0.45\textwidth}
    \includegraphics[width=\textwidth]{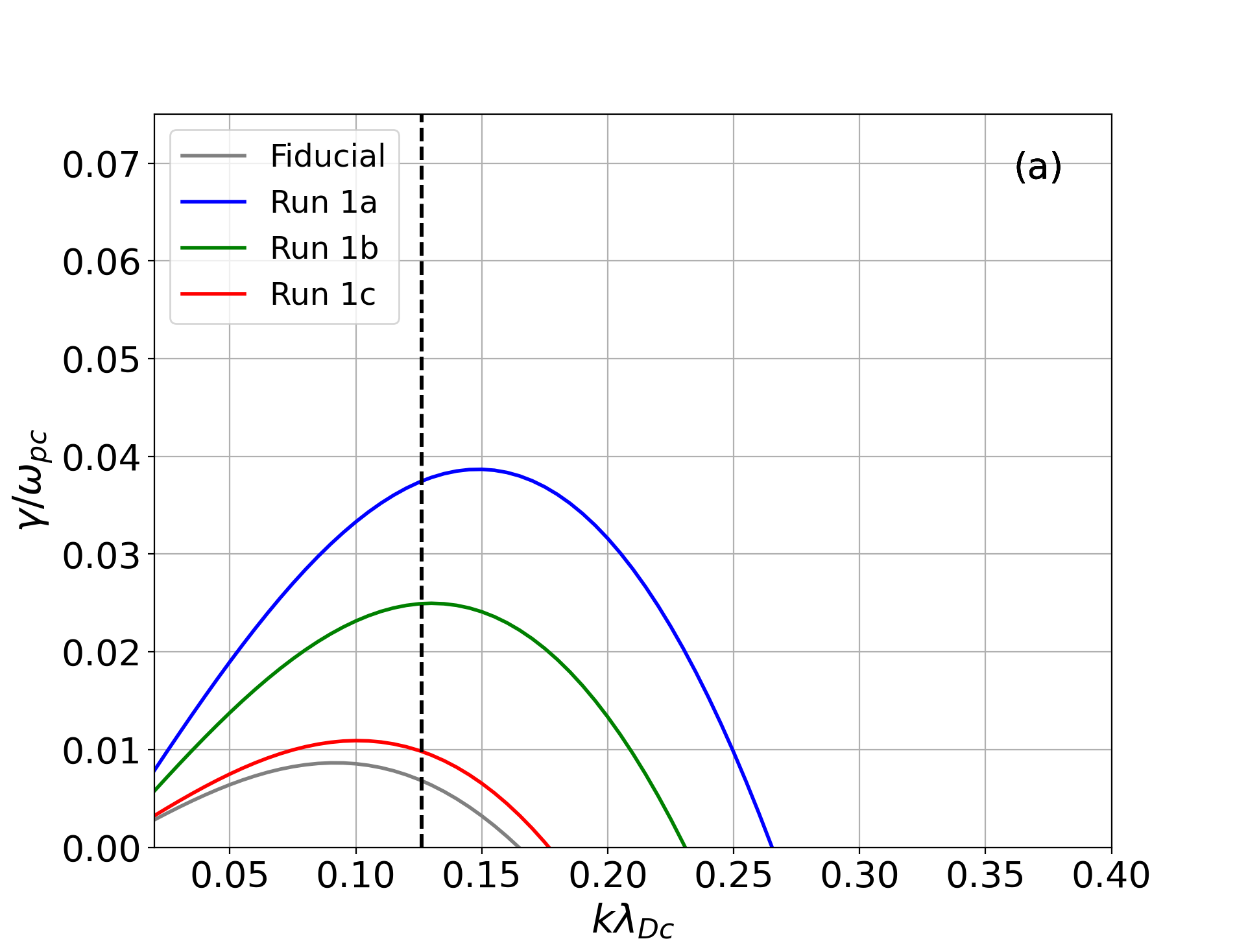}
\end{minipage}
\begin{minipage}[b]{0.45\textwidth}
    \includegraphics[width=\textwidth]{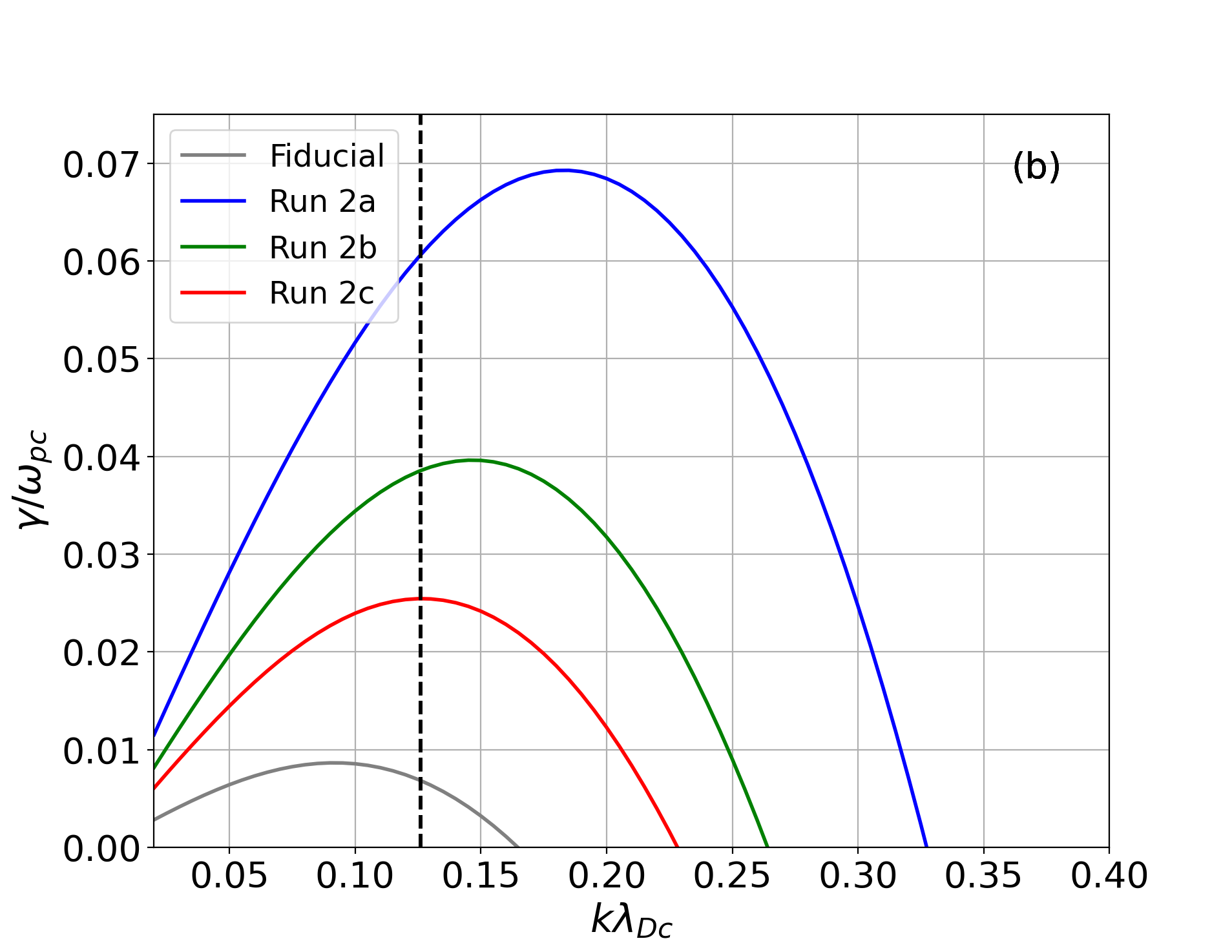}
    \end{minipage}
\begin{minipage}[b]{0.45\textwidth}
    \includegraphics[width=\textwidth]{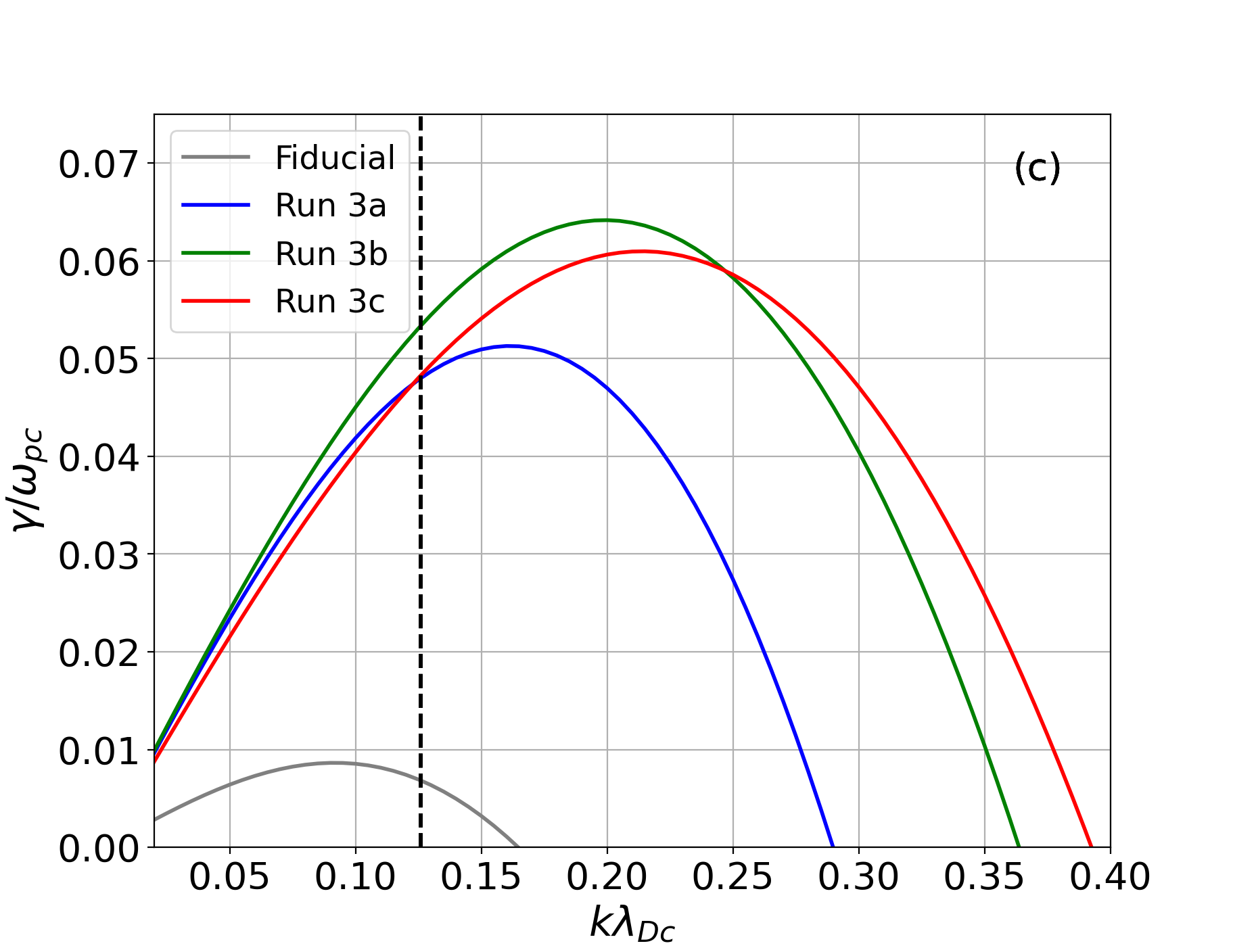}
\end{minipage}
\caption{Normalized growth rate, $\gamma/\omega_{pc}$ vs. normalized wavenumber $k \lambda_{Dc}$ from Eq. \ref{DR} for the IIAI with parameters given by Table~\ref{tab:runs}. Panel a, b, and c depict Series 1, 2 and 3, respectively, with the fiducial run shown by the gray curve. These theoretical expectations are compared with simulation results in Table \ref{comparison}. The vertical dashed lines refer to the simulated wavenumber ($k\lambda_{Dc}=0.126$).}
\label{fig:DR}
\end{figure}

\begin{figure}[h!]
\centering
\begin{minipage}[b]{0.45\textwidth}
    \includegraphics[width=\textwidth]{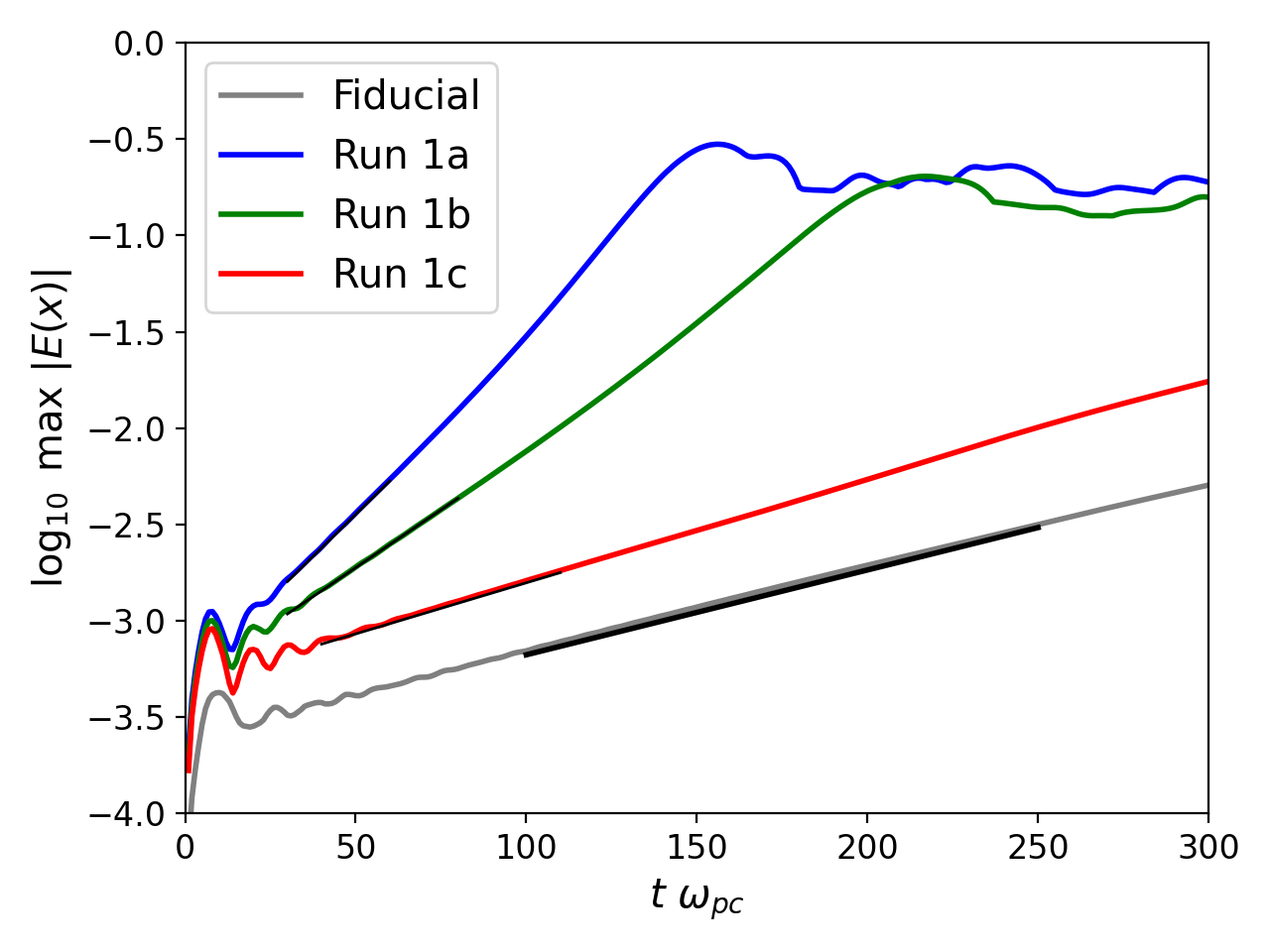}
\end{minipage}
\begin{minipage}[b]{0.45\textwidth}
    \includegraphics[width=\textwidth]{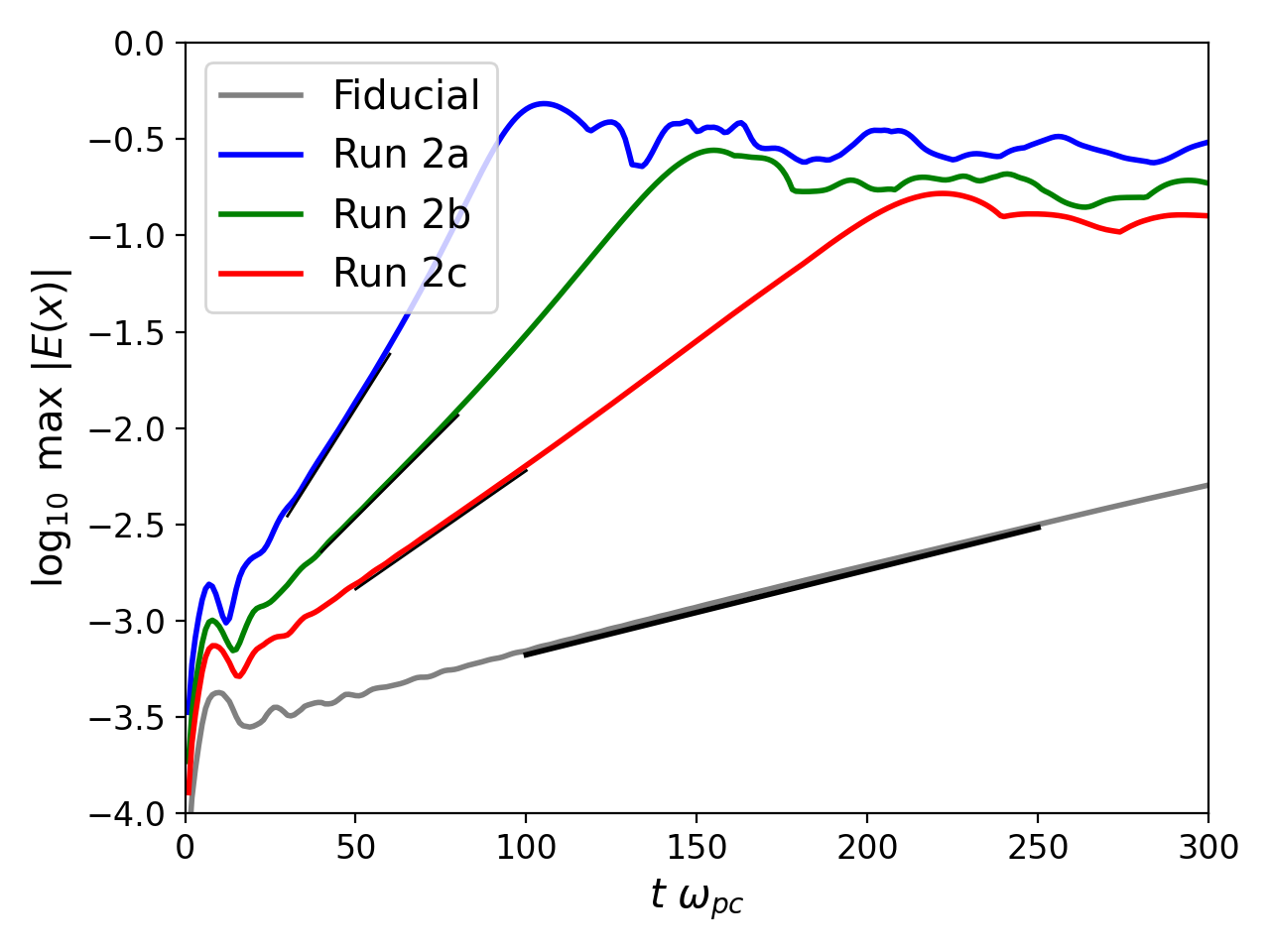}
\end{minipage}
\begin{minipage}[b]{0.45\textwidth}
    \includegraphics[width=\textwidth]{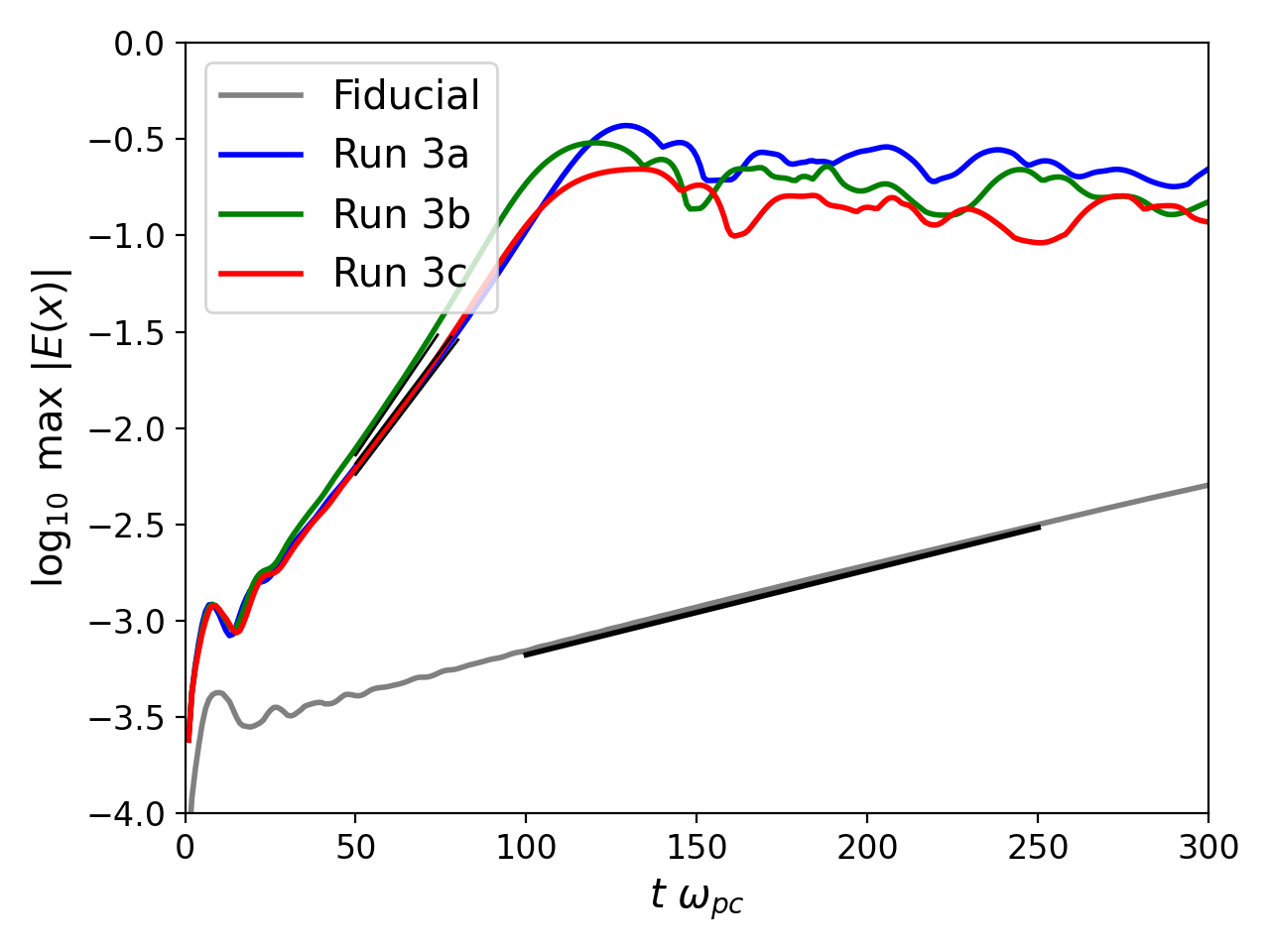}
\end{minipage}
\caption{Evolution of the electric field amplitude as a function of time for the simulation runs described in Table~\ref{tab:runs}. 
In Table~\ref{comparison}, we compare the growth rate measured here, illustrated with black lines, with linear theory.}
\label{fig:sim_runs}
\end{figure}

The growth rate's dependence on key system parameters is evident in Figs.~\ref{fig:DR} and~\ref{fig:sim_runs}, where we show results from IIAI linear theory and simulations, respectively, using the parameters listed in Table~\ref{tab:runs}. 
 The simulation parameters in the table integrate our previous dilute-beam results \citep{Afify2024} with new regimes motivated by in situ measurements of nonthermal ion populations \citep{Verniero2020, Verniero2022, Phan2022}. 
 The Fiducial run maintains the core-beam configuration from \citet{Afify2024} but adopts an enhanced beam-core drift $V_{D,b}/v_{th,c} = 5.7$, consistent with PSP observations of the beam proton \citep{mozer2021triggered}, where drifts of $\sim180$ km/s far exceeded the ion thermal speed of $\sim 27 $ km/s.
 Series 1–3 systematically probe critical parameter dependencies: Series 1 varies $T_e/T_c$ to match the elevated electron-to-core temperature ratios ($7 \lesssim T_e/T_c \lesssim 10$) measured during intervals of IAW activity \citep{Pisa2021, Mozer2022}, Series 2 explores $n_b/n_c$ values ($0.1$–$0.3$) spanning the transition from dilute beams in quiet solar wind to dense beams in magnetic reconnection regions \citep{Phan2022}, while Series 3 probes $V_{D,b}/v_{th,c}$ variations compatible with observations in~\citet{Graham2021}. 
 This parameter space directly samples the conditions underlying the bursty IAWs reported in these studies \citep{mozer2021triggered, Pisa2021, Graham2021, Mozer2022}, enabling quantitative comparison between our predicted instability threshold and heating rates and observed nonthermal proton distributions. 

In Fig.~\ref{fig:DR}, panels a, b, and c, we depict Series 1, 2 and 3, respectively; the fiducial run is represented by the gray curve. 
We observe a strong enhancement in growth rate with increasing electron-to-core temperature ratio (panel a) and beam-to-core density ratio (panel b). 
In panel c we observe the same pattern relating increasing drift velocity and growth rate already described in~\citet{Afify2024}, Fig. 1: the growth rate first increases and then decreases with increasing core-to-beam drift.

These (already well known) theoretical results are confirmed by simulations.
The simulations are run using a 1D1V Vlasov simulation framework \citep{Afify2024, Afify2025}. 
The simulation boxes have size $L_x/ \lambda_{Dc}=50$, which selects the wavenumber $k \lambda_{Dc}=0.126$ if only one oscillation is present in the box. 
These values were selected to ensure the simulations resolve a wavenumber close enough to that of maximum growth, according to linear theory analysis.
The solver implements a Finite-Volume method to discretize the Vlasov equation on a two-dimensional coordinate-velocity phase-space grid. 
The evolution of the cell-averaged values of the distribution function, $f_s$, is performed using a third-order Runge-Kutta scheme \citep{SHU1988439}. 
Periodic boundary conditions are applied along the $x$-axis, while zero-flux boundary conditions are imposed along the $v$-axis, ensuring the conservation of the total particle number within the domain for each species, except for the occasional resetting of negative undershoots in $f_s^\alpha$ to zero. 
The integration time step is chosen as $\Delta t = 1/2500 \omega_{pc}$, as in~\citet{Afify2024}.   
Initial particle distribution functions are modeled as drifting Maxwellians according to Eq. \ref{equation3}. 
To excite the fundamental mode with parameters given by Table \ref{tab:runs}, a perturbation  $\delta V= 0.01 \sin \left( \left(2 \pi/ L_x \right) x \right)$ is added to the beam drift velocity.

In Fig.~\ref{fig:sim_runs} we show the evolution as a function of time of the maximum electric field value for Series 1, 2 and 3 runs in panels a, b, and c, respectively, where the fiducial simulation is presented by the gray curve. 
From these plots, we calculate the growth at the wavenumber  $k\,\lambda_{Dc}=0.126$ selected by the simulation box length and we compare it, together with the real frequency, with results from linear theory in Table \ref{comparison}. We observe good agreement between theoretical and simulated results.

\begin{table}[]
    \centering
    \caption{Key parameters of the runs under consideration. $,n_b/n_c$, $T_b/T_c$, and $T_e/T_c$ denote the beam-to-core density ratio, beam-to-core temperature ratio, and electron-to-proton core temperature ratio, respectively, while $V_{D,b}/v_{th,c}$ and $V_{D,e}/v_{th,c}$ are the drift velocities of the proton beam and electrons normalized to the proton core thermal speed. In Series 1, 2, and 3, we reduce $T_e/T_c$, $n_b/n_c$, and $V_{D,b}/v_{th,c}$, respectively; unchanged parameters are indicated by “–”.}
    \begin{tabular}{|c|ccccc|}
    \hline
         Run & $\frac{n_b}{n_c}$ & $\frac{T_b}{T_c}$ & $\frac{T_e}{T_c}$ & $\frac{V_{D,b}}{v_{th,c}}$ & $\frac{V_{D,e}}{v_{th,c}}$  \\
    \hline
         Fiducial & 0.05 & 1.0 & 10.0 & 5.7 & 0.285 \\
         \hline
        1a & 0.2 & 1.0 & 9.0 & 5.7 & 0.95 \\ 
         1b & - & - & 8.0 & - & - \\ 
         1c & - & - & 7.0 & - & - \\ 
    \hline
         2a & 0.3 & 1.0 & 10.0 & 5.7 & 1.32\\ 
         2b & 0.15 & - & - & - & 0.74\\ 
         2c & 0.1 & - & - & - & 0.52 \\ 
    \hline
         3a & 0.2 & 1.0 & 10.0 & 5.7 & 0.95 \\ 
         3b & - & - & - & 5.0 & 0.83 \\ 
         3c & - & - & - & 4.5 & 0.75 \\ 
    \hline
    \end{tabular}
    \label{tab:runs}
\end{table}

\begin{table}[]
\centering
\caption{Comparison of linear theory and simulation results for the simulations listed in Table~\ref{tab:runs}. Both theoretical and simulated results are for wavenumber $k \lambda_{Dc}= 0.126$.}
\label{comparison}
\begin{tabular}{|c|c|c|c|c|}
\hline
\multirow{2}{*}{Run} & 
\multicolumn{3}{c|}{Linear Theory} & 
\multicolumn{1}{c|}{Simulation} \\
\cline{2-5}
& $\frac{\omega}{\omega_{pc}}$ & $\frac{\gamma}{\omega_{pc}}$ &   $\frac{v_{res}}{v_{th,c}}$ & $\frac{\gamma}{\omega_{pc}}$  \\
\hline
Fiducial & 0.42 & 0.0068  & 3.333 & 0.01 \\
\hline
1a & 0.309 & 0.0375  & 2.452 & 0.04  \\
\hline
1b & 0.301 & 0.0249  & 2.389 & 0.028  \\
\hline
1c & 0.292 & 0.00982  & 2.317 & 0.012  \\
\hline

2a & 0.255 & 0.061  & 2.024 & 0.0645  \\
\hline
2b & 0.349 & 0.0385  & 2.77 & 0.041  \\
\hline
2c & 0.385 & 0.0254  & 3.06 & 0.0283  \\
\hline

3a & 0.316 & 0.048  & 2.508 & 0.0537  \\
\hline
3b & 0.296 & 0.053  & 2.349 & 0.0599  \\
\hline
3c & 0.282 & 0.0483  & 2.238 & 0.0541  \\
\hline

\end{tabular}
\vspace{-1em}
\end{table}

\section{Energy Evolution}
\label{ssec:method.simulation}
\begin{figure}[h!]
\centering
\begin{minipage}[b]{0.4\textwidth}
    \centering
    \includegraphics[width=\textwidth]{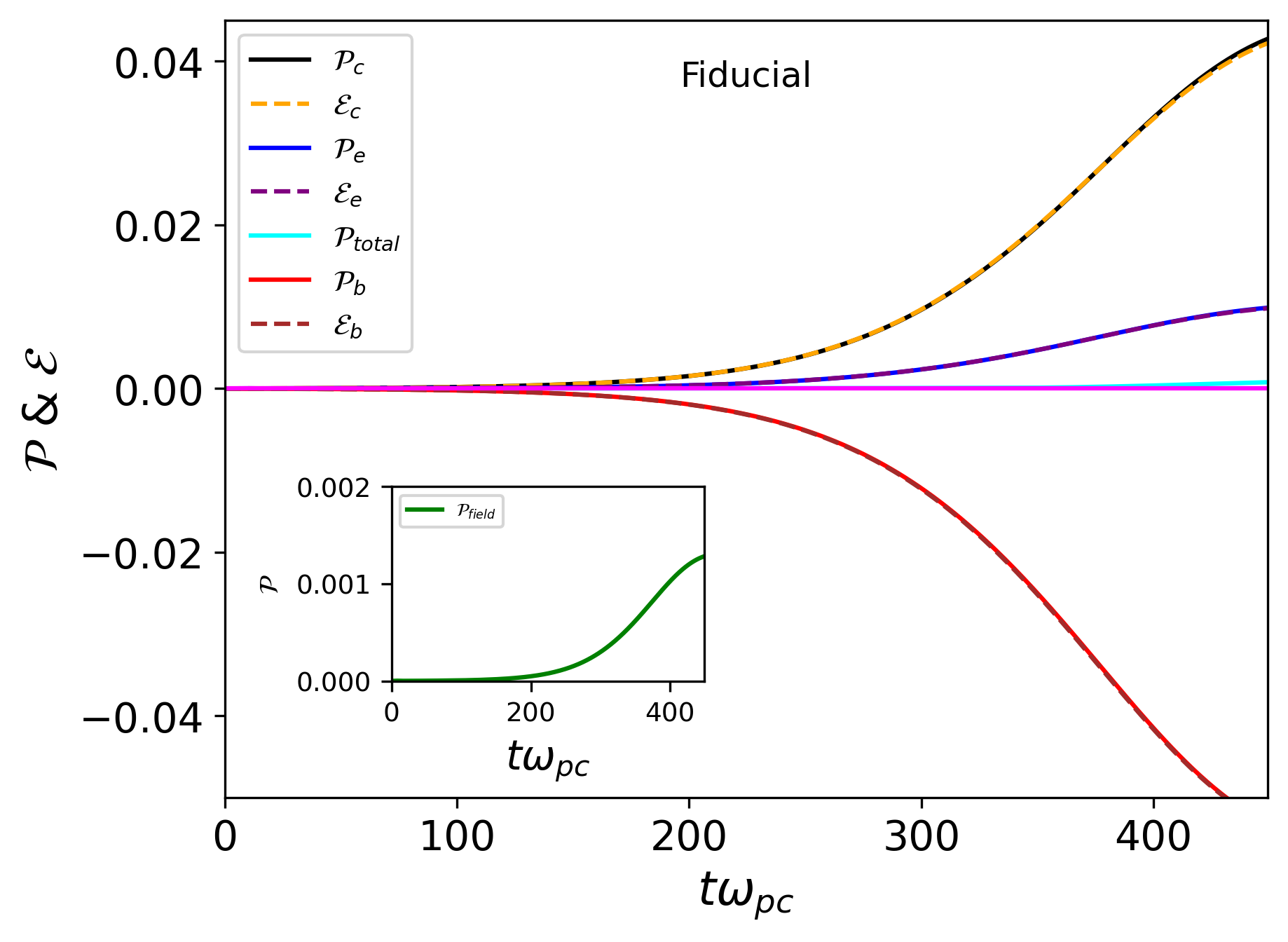}
\end{minipage}
\begin{minipage}[b]{0.4\textwidth}
    \centering
    \includegraphics[width=\textwidth]{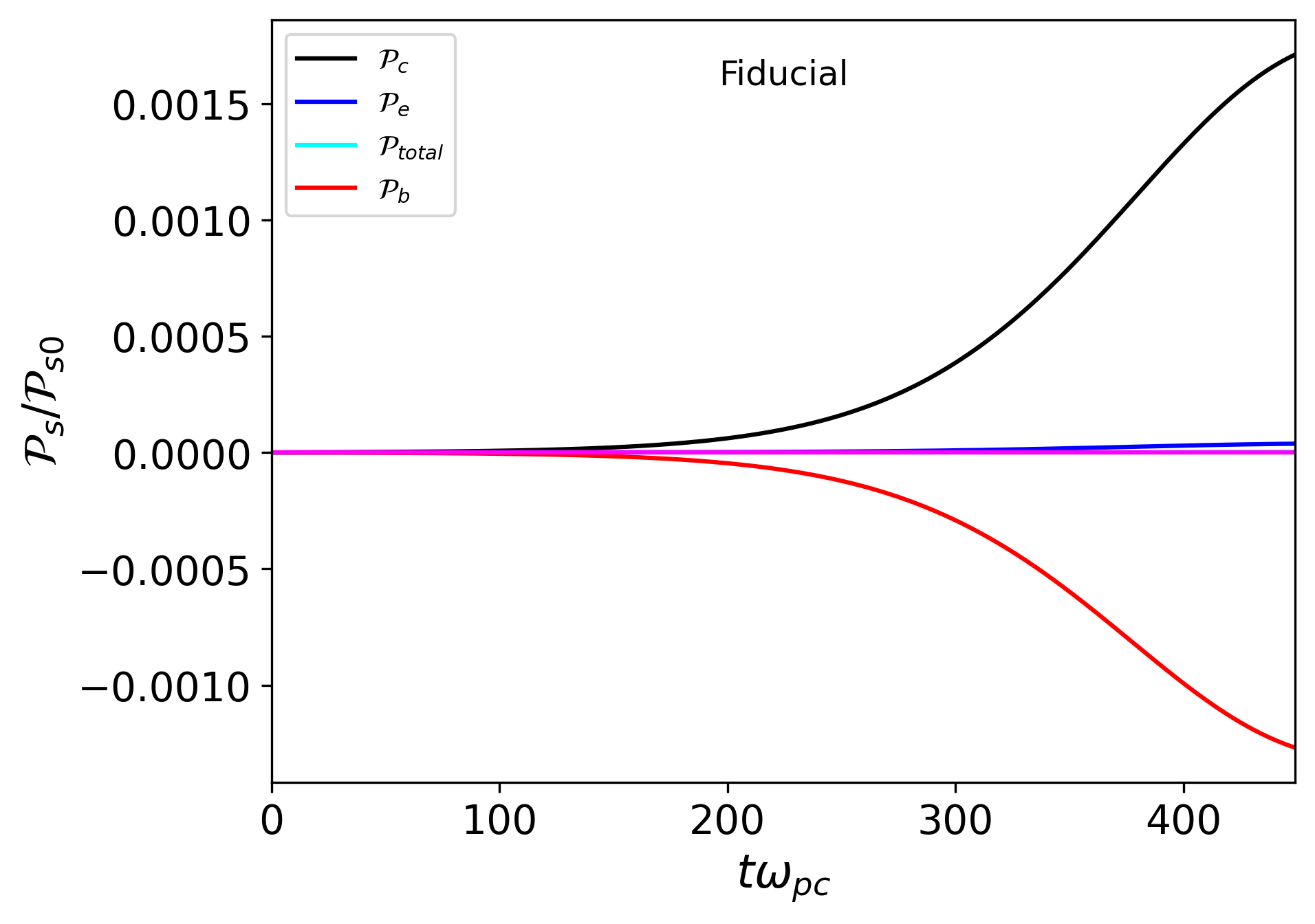}
\end{minipage}
\vspace{0.5em}
\caption{Energy evolution in the fiducial run.
Upper panel: (Solid lines) energy variation of the different particle populations, the total energy, and of the electric field energy (inset) calculated as Eq.~\ref{normal}. (Dashed lines): energy variation for the particle populations calculated as Eq.~\ref{barw}. 
Lower panel: $\mathbf{\mathcal{P}_s/ \mathcal{P}_{s0}}$ for the three particle populations and for the total energy. The horizontal solid magenta line at zero separates positive and negative values of $\mathbf{\mathcal{P}}$ and $\mathbf{\mathcal{E}}$.}
\label{fig:Eng_Fiducial}
\end{figure}

\begin{figure*}[h!]
\centering
\begin{minipage}[b]{0.3\textwidth}
    \includegraphics[width=\textwidth]{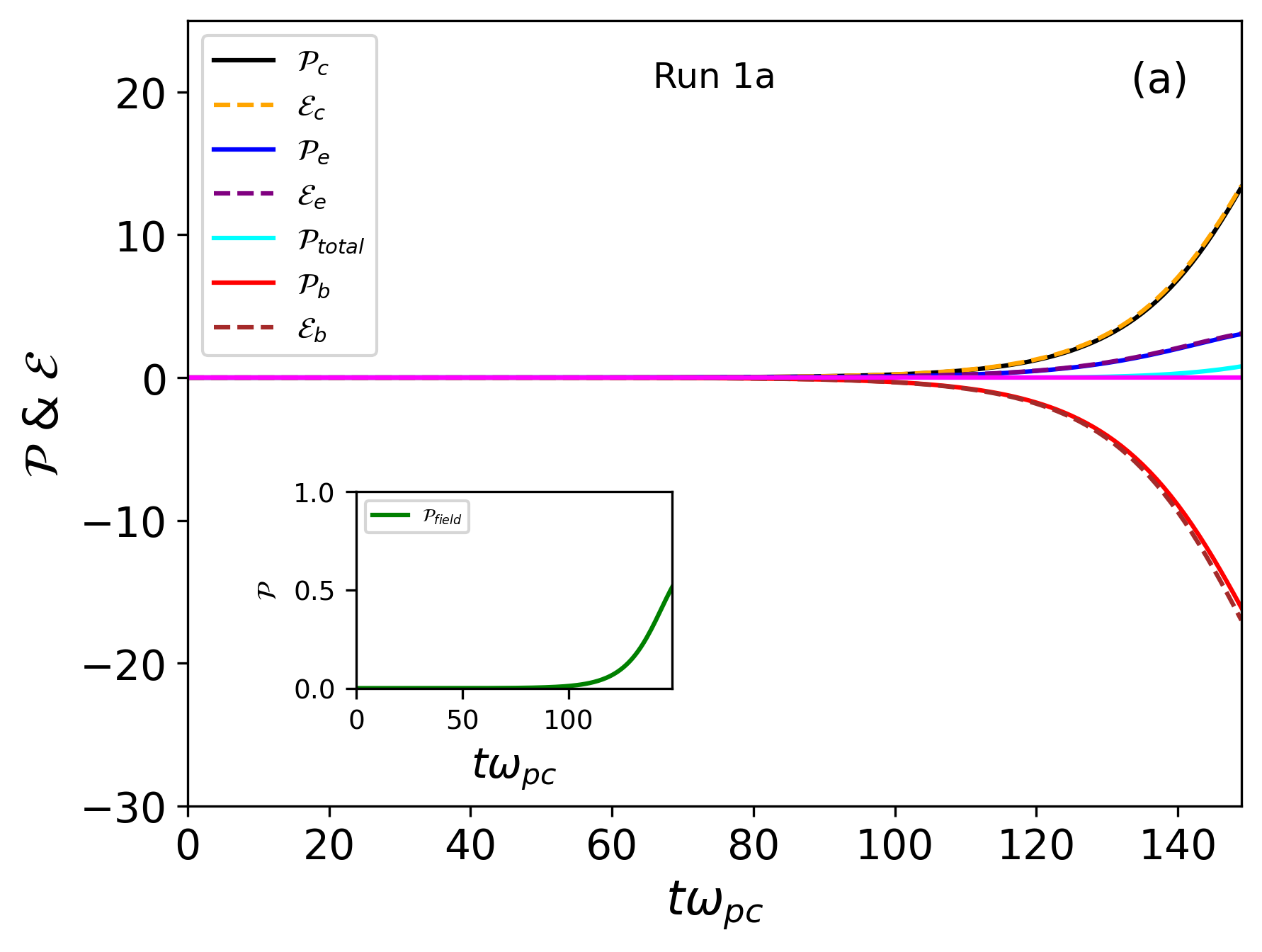}
\end{minipage}
\begin{minipage}[b]{0.3\textwidth}
    \includegraphics[width=\textwidth]{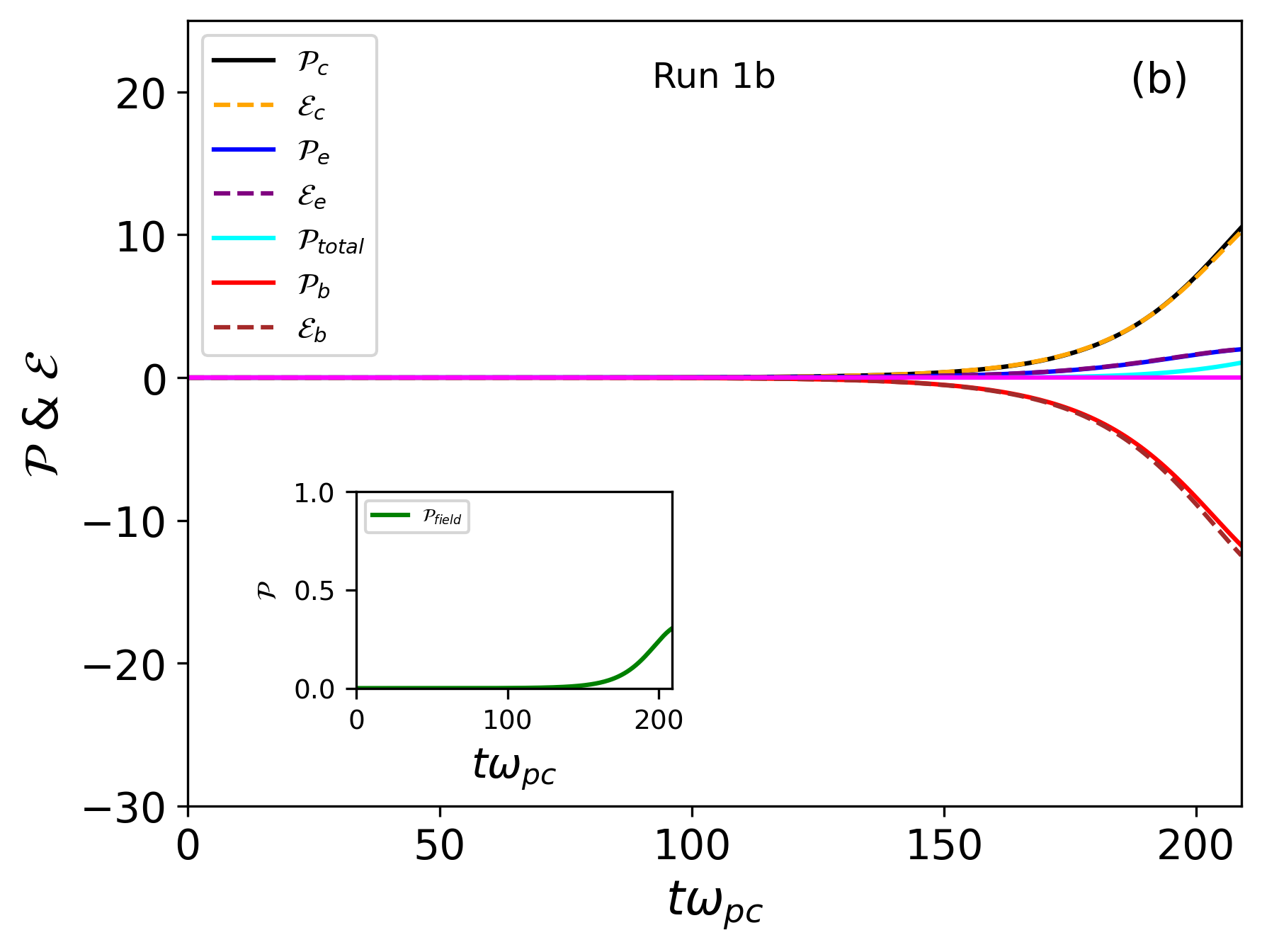}
\end{minipage}
\begin{minipage}[b]{0.3\textwidth}
    \includegraphics[width=\textwidth]{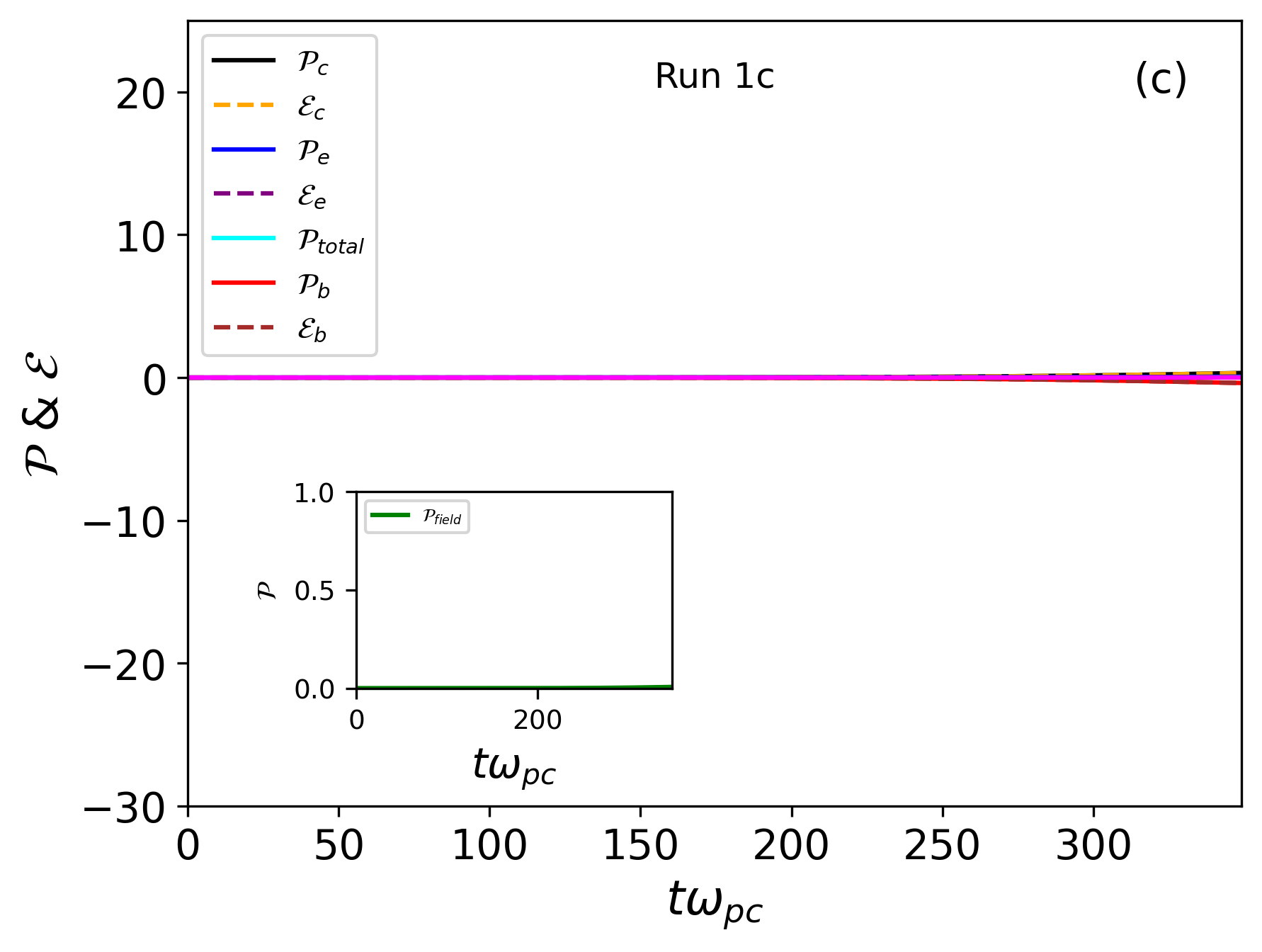}
\end{minipage}
\begin{minipage}[b]{0.3\textwidth}
    \includegraphics[width=\textwidth]{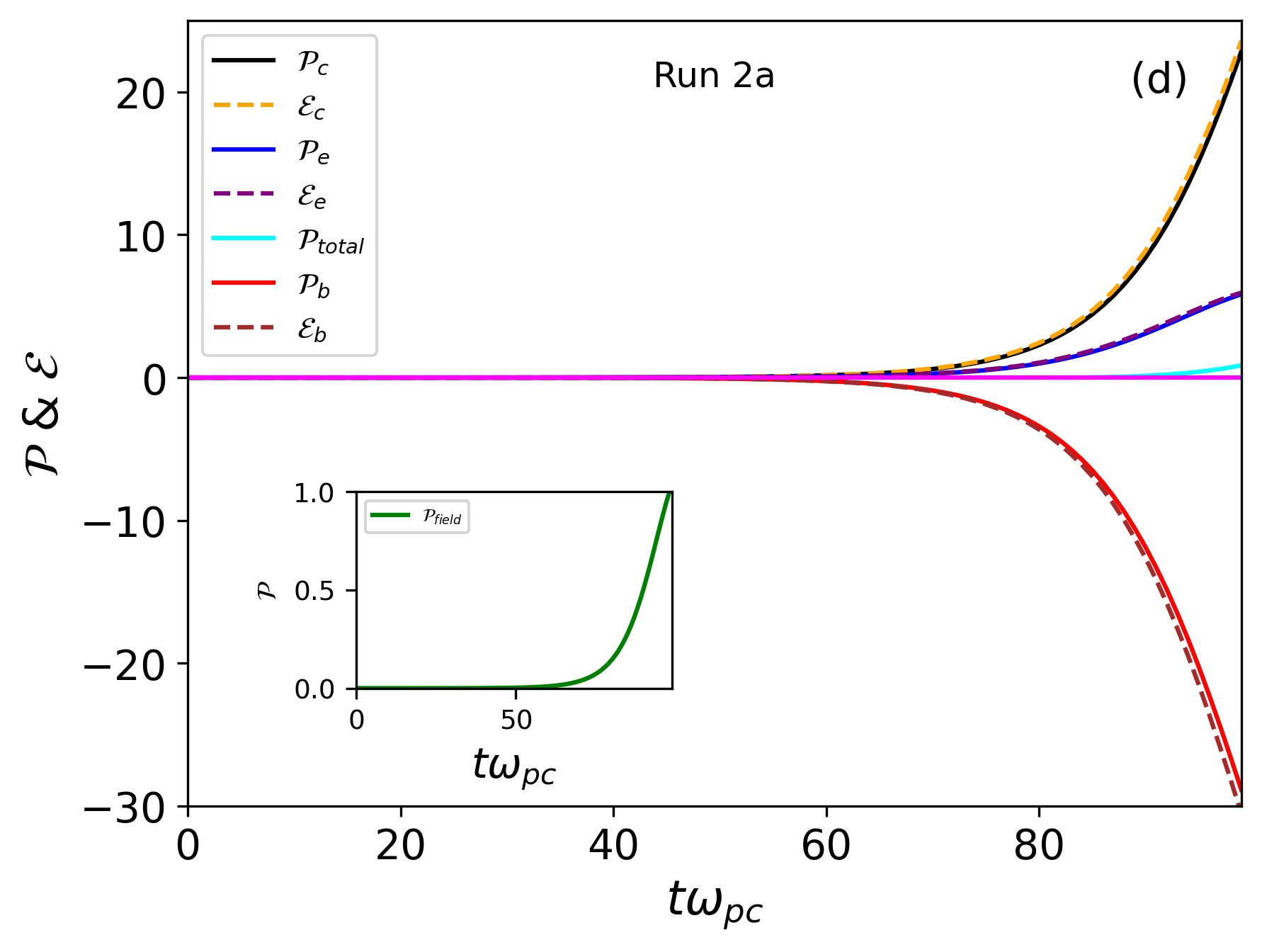}
\end{minipage}
\begin{minipage}[b]{0.3\textwidth}
    \includegraphics[width=\textwidth]{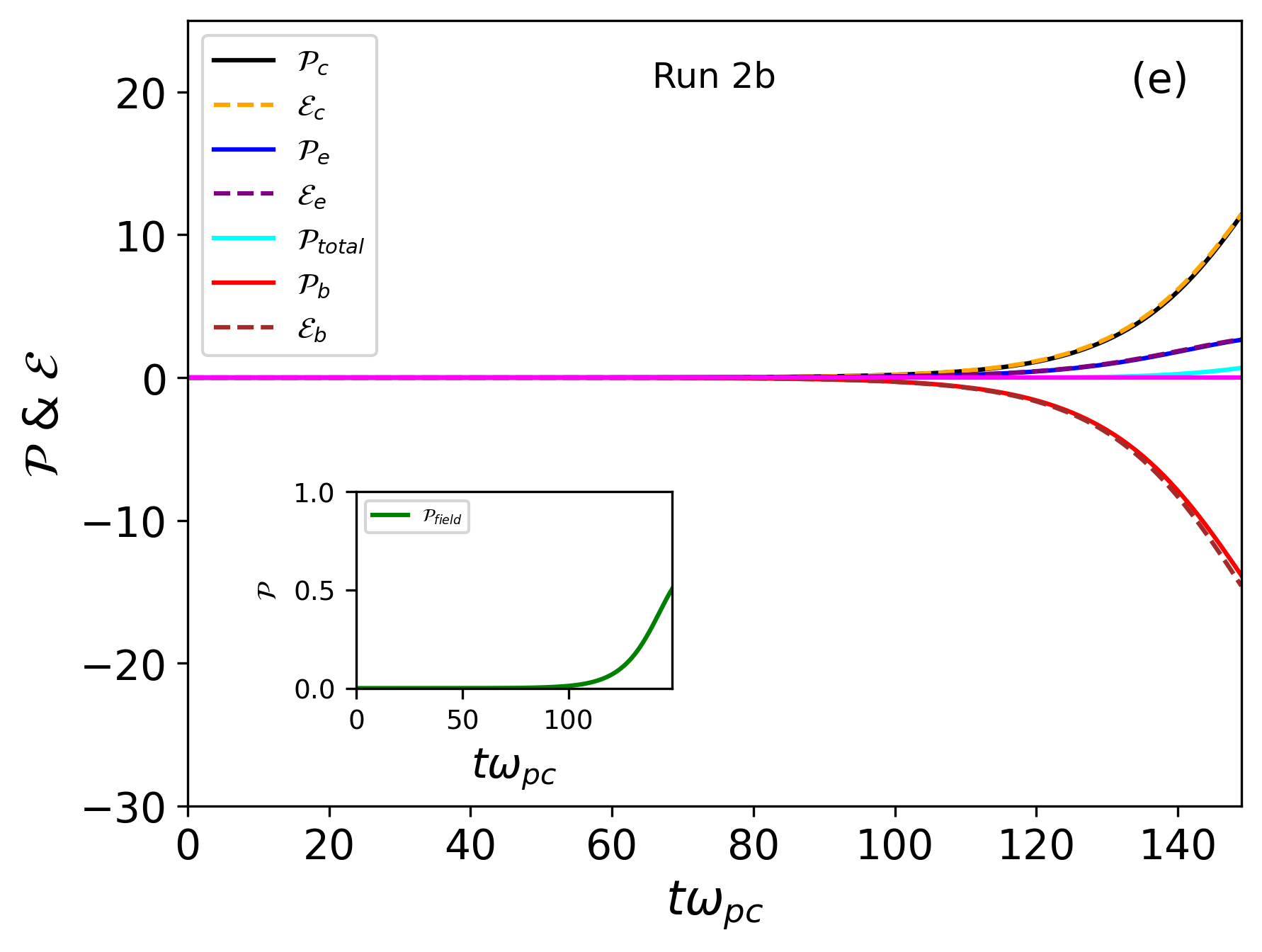}
\end{minipage}
\begin{minipage}[b]{0.3\textwidth}
    \includegraphics[width=\textwidth]{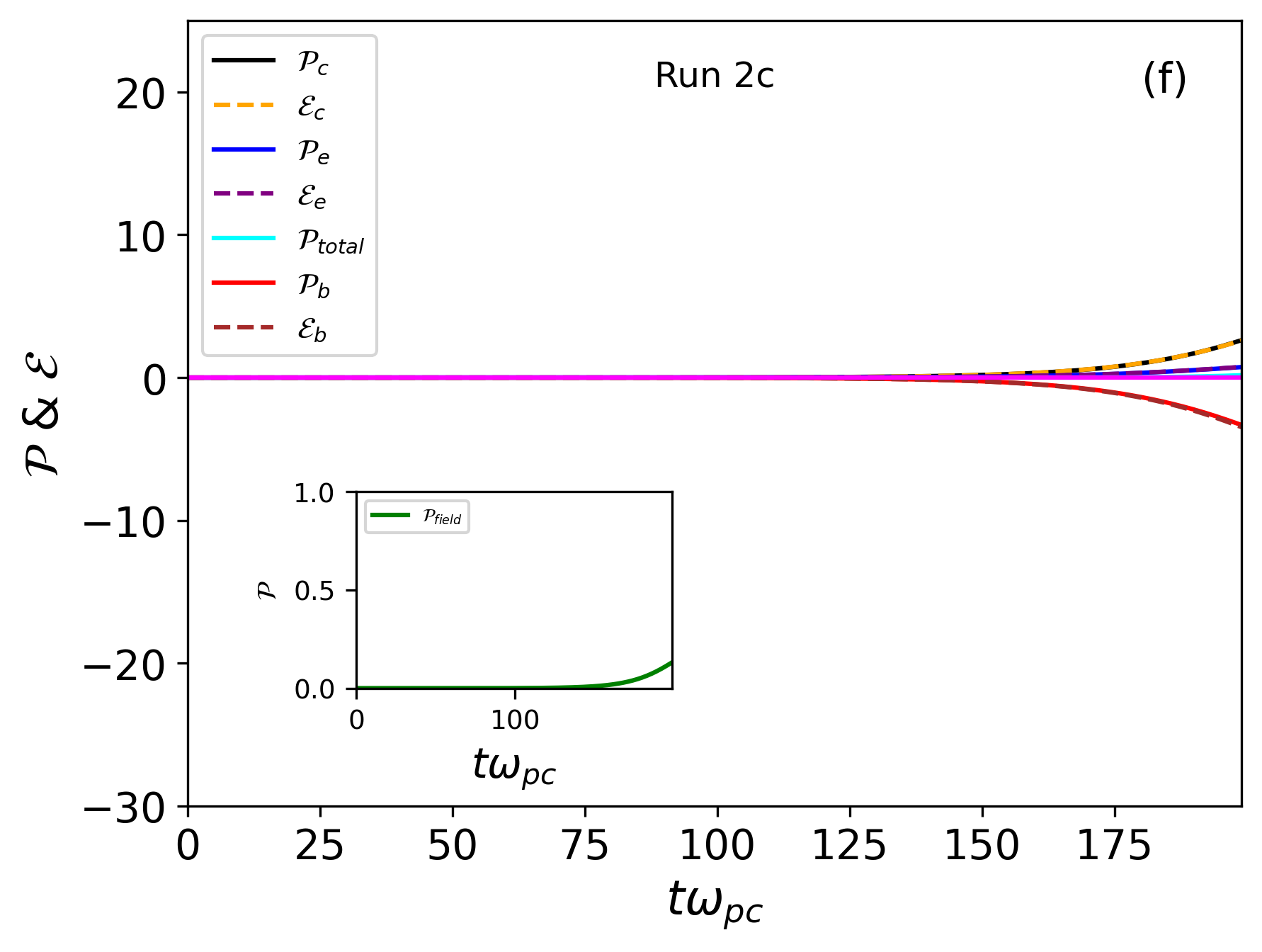}
\end{minipage}
\begin{minipage}[b]{0.3\textwidth}
    \includegraphics[width=\textwidth]{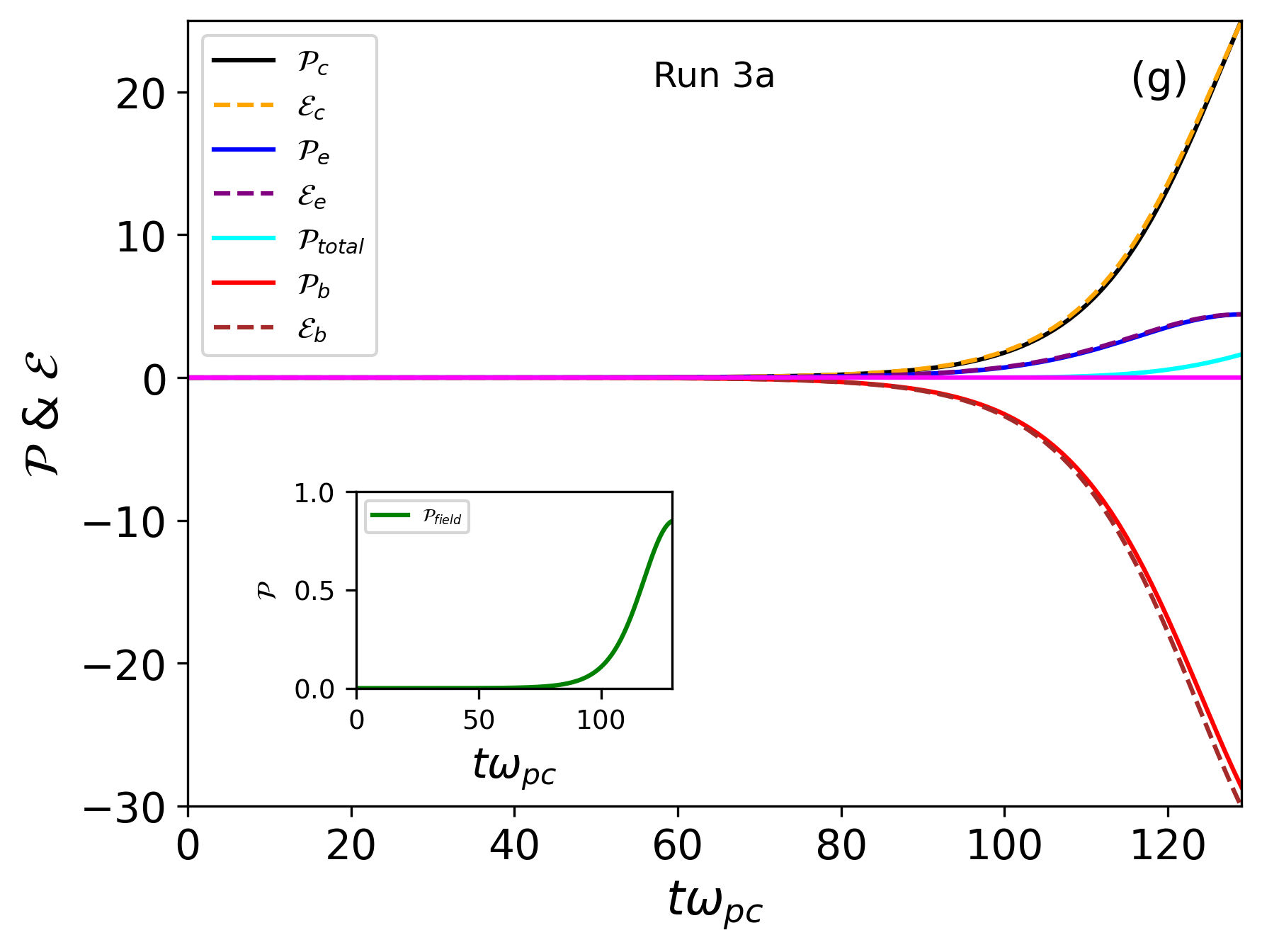}
\end{minipage}
\begin{minipage}[b]{0.3\textwidth}
    \includegraphics[width=\textwidth]{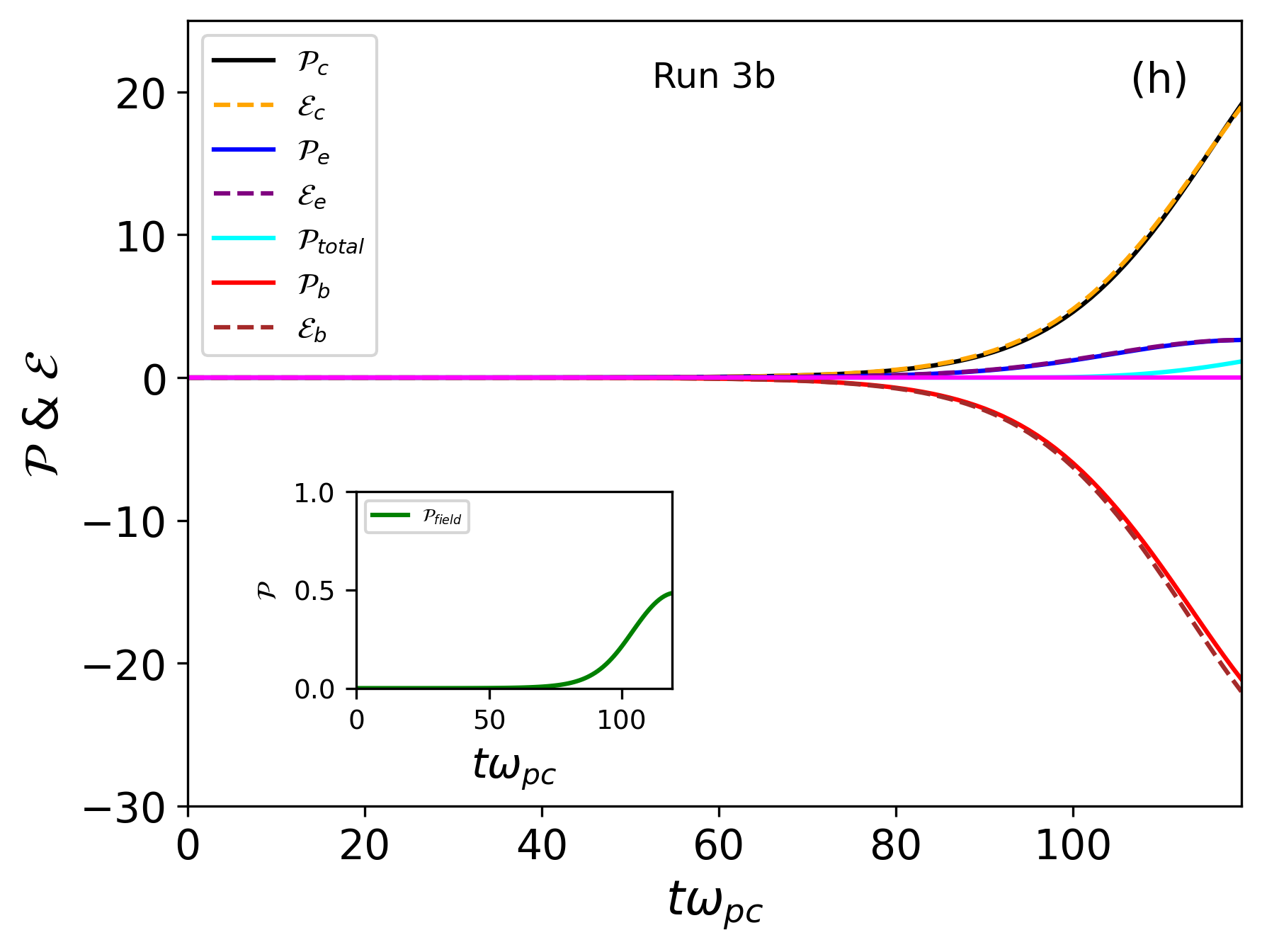}
\end{minipage}
\begin{minipage}[b]{0.3\textwidth}
    \includegraphics[width=\textwidth]{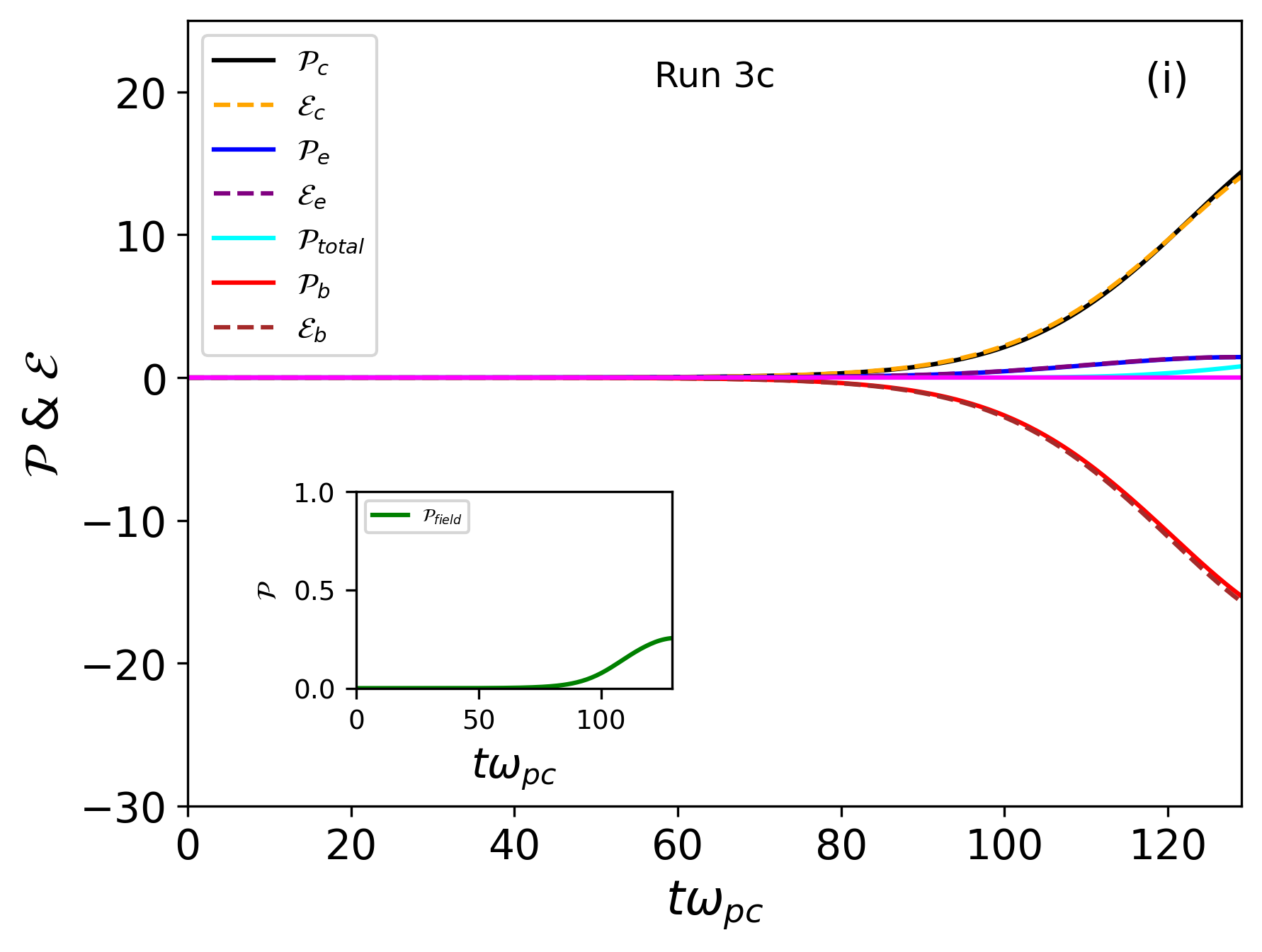}
\end{minipage}
\caption{The change in the particles, field, and total kinetic energy (solid lines,$\mathcal{P}$), and the net energy exchange rate (dashed lines,$\mathcal{E}$),  for Series 1 (first row), Series 2 (second row), and Series 3 (third row) runs in Table~\ref{tab:runs}. Same as in Fig.~\ref{fig:Eng_Fiducial}, with the horizontal zero line indicating the separation between positive and negative values.}
\label{fig:Eng_runs}
\end{figure*}

In this Section we focus on energy exchange between fields and particles in our simulations.
Multiplying Eq. \ref{eq:vlasov} by $\frac{1}{2}m_s v^2$ and then integrating over all coordinate (1D) and all velocity (1V) space yields: 
\begin{multline}
\int d x \int d v \frac{m_s v^2}{2} \frac{\partial f_s}{\partial t}+\int d x \int d v  \frac{m_s v^2}{2} v \frac{\partial f_s}{\partial x}\\
-\int d x \int d v  \frac{m_s v^2}{2} \frac{q_s}{m_s} \frac{\partial \varphi}{\partial x} \frac{\partial f_s}{\partial v}=0.
\label{eq:modified poisson}
\end{multline}
The second term in Eq.~\ref{eq:modified poisson} vanishes for periodic boundary conditions and for distributions that decay at infinity, since it contributes only through boundary terms. We define the phase-space energy density for species $s$ as $\epsilon_s(x,v,t) =   \frac{m_s v^2}{2} f_s(x,v,t)$, which represents the energy density per unit length and per unit velocity in the 1D–1V Vlasov–Poisson system \citep{howes2017diagnosing}. The time evolution of the phase-space energy density, $\partial \epsilon_s(x,v,t)/\partial t$, can then be obtained from Eq.~\ref{eq:modified poisson}, yielding
\begin{equation}
\frac{\partial \epsilon_s(x, v, t)}{\partial t} = -\frac{q_s v^2}{2} \frac{\partial f_s(v)}{\partial v} E(x, t), \label{eq:secularEnergy}
\end{equation}
Therefore, Eq. \ref{eq:secularEnergy} gives the instantaneous rate of energy transfer between the electric fields and the particle species, which is responsible for variation in particle energy.
Integrating Eq.~\ref{eq:secularEnergy} over space, velocity, and time, one gets
\begin{multline}
\mathcal{E}_s = -\frac{1}{2}\int_0^t dt^{\prime} \int dx \int dv \, q_s v^2 \\
\times \left(\frac{\partial  f_s(x, v, t^{\prime})}{\partial v}\right) E(x, t^{\prime}).
\label{barw}
\end{multline}
i.e. the non-linear wave particle interaction term.
The evolution of the electric field energy $W_{\varphi} = \int \mathrm{d}x \frac{E^2}{2}$ and the microscopic particle energy $W_s = \int \mathrm{d}x \int \mathrm{d}v  \frac{m_s v^2}{2} f_s$ is shown in Fig.~\ref{fig:Eng_Fiducial} (panel a) for the fiducial simulation and in Fig.~\ref{fig:Eng_runs} for the other simulations in Table~\ref{tab:runs}. The solid lines depict 
\begin{equation}
\mathcal{P}_s = W_s - W_{s0},
\label{normal}
\end{equation}    
where $s$ is the core, beam, electron, field and total energy, and the subscript $0$ labels the initial time. These values are obtained directly from usual simulation diagnostics. 
The dashed lines similarly depict $\mathcal{E}_s$ for the different particle populations as per Eq.~\ref{barw}.

We observe in Fig.~\ref{fig:Eng_Fiducial}, panel a and in Fig.\ref{fig:Eng_runs} that the solid and the dashed lines superimpose for the particle populations, as they should since they represent different ways of calculating energy exchange for a particle species.
The cyan curves track the total energy changes in the simulation (fields plus particles), illustrating excellent energy conservation in our runs. 
We see in Fig.~\ref{fig:Eng_Fiducial} that the main energy exchange is between the core and beam population: the beam loses energy, which is gained mostly by the core protons and, to a lower extent, by the electrons. The energy gained by the electric field is small in comparison, as the field acts primarily as a medium for energy transport between the particle species. 
In Fig.~\ref{fig:Eng_Fiducial}, panel b, we plot the energy variation of protons (core and beam) and of the electrons normalized to their respective initial energy, $\mathcal{P}_s/ \mathcal{P}_{s0}$: we see that the relative energy variation of the electron is minimal with respect to the one of the core and beam protons: the electrons do not gain significant kinetic energy.

In Fig.~\ref{fig:Eng_runs} we see that the amount of exchanged energy depends directly on the growth rate of the instability, as expected: higher growth rate is accompanied by a larger amount of exchanged energy. In the first line of Fig.~\ref{fig:Eng_runs} decreasing $T_e/T_c$ results in decreasing growth rate (Table~\ref{comparison}): the $\mathcal{P}$ components decrease in absolute value. In the second line of Fig.~\ref{fig:Eng_runs}, decreasing $n_b/n_c$ also results in lower growth rates (Table~\ref{comparison}): again, the amount of energy exchanged between the population decreases. In the third line of Fig.~\ref{fig:Eng_runs}, we see minimal difference in the amount of energy exchange as a function of increasing $V_{D,b}/ v_{th,c}$, with respect to the larger variation observed in the first and the second row. The reason for this is clear from Fig.~\ref{fig:sim_runs}: the variation in growth rate between the three simulations is much smaller in Series 3 than in Series 1 and 2. 
The field energy evolution is depicted in the inset panels, and it exhibits the same pattern: weaker instabilities result in a smaller field energy increase and, consequently, weaker particle energization.
Moreover, a decrease in $T_e/T_c$, $n_b/n_c$, or $V_{D,b}/v_{th,c}$ in all three parameter scans leads to a reduced electron energy gain relative to the beam's energy loss. This suggests that when instability becomes less effective, electrons absorb a lesser percentage of the available energy, which results in a slight shift in the energy partition between the electrons and the beam, along with a decrease in the total energy exchanged.

\section{Field Particle Correlation}
\label{sec:FPC}
Investigation of the evolution of VDFs and electromagnetic fields is constrained by the limitations of in situ observations, which typically rely on single-point measurements. 
Consequently, the nonlinear wave-particle interaction described by Eq.~\ref{eq:secularEnergy} presents a tailored methodology for examining the rate of change of phase-space energy density, as opposed to total energy, leveraging single-point time-series measurements of electromagnetic fields and VDFs. 
The field-particle correlation (FPC) technique quantifies energy transfer by correlating the electric field (which alone performs work) with the velocity derivative of the VDF. 
This single-point measurement approach reveals the velocity-space structure of the field-particle energy exchange which can be tied to specific energization and instability processes.
The FPC approach is fundamentally applicable to single-point spacecraft measurements, providing a powerful tool for analyzing particle energization mechanisms in weakly collisional heliospheric plasmas.

To differentiate between oscillatory energy exchange, associated with wave motion, and secular energy exchange resulting from net energy transfer between particles and fields, the FPC method performs a time average over an interval $\tau = N \Delta t$ exceeding the linear wave period (i.e., $\tau > \frac{2\pi}{\omega}$ for the dominant oscillations). This isolates the secular component of the energy transfer, while the oscillatory component, representing reversible energy exchange, is eliminated.
The field-particle correlation at time \( t_i \) and position \( x_0 \), computed over an interval \( \tau = N \Delta t \), is given by \citep{Klein2017a}:
\begin{align}
\label{eq:FPC}
FPC(x_0, v, t_i, \tau) \equiv \frac{1}{N} \sum_{j=i}^{i+N} 
    & -q_s \frac{v^2}{2} 
    \frac{\partial  f_{s,j}(x_0, v, t_j)}{\partial v} \notag \\
    & \times E(x_0, t_j).
\end{align}
where the term that enters the summation is the right hand side of Eq.~\ref{eq:secularEnergy} evaluated at a single point in space, $x=x_0$.
By integrating over all spatial positions, we can compare FPC to the variation of species energy in Eq.~\ref{barw} at a given time $t'$.
By analyzing single-point measurements at position $x= x_0$ we can obtain a map of FPC for each particle species $s$, as a function of velocity and time, from which we can directly see at which velocities energy is exchanged by that species with the electric field, if the exchange is secular or the result of oscillations, and if the particle species gains (positive sign) or loses (negative sign) energy.

In the following subsections, we examine FPC as a function of velocity and integrated over velocity, $\int dv FPC(x_0, v, t_i, \tau)$,
looking for phase-space signatures at a specific spatial location for each simulation run listed in Table \ref{tab:runs}.

\subsection{Fiducial Run}
\label{ssec:results.fiducial}
In Fig.~\ref{fig:fiducial_fpc} we plot  $\int dv FPC $, which quantify the interaction between the electric field $E$ and the three plasma populations over correlation intervals $\tau \omega_{pc} \in [0, 100]$. The quantity is depicted in normalized units, which follow from the code normalization described in Sec.~\ref{sec:linear-analysis}. 
With the real frequencies of our instabilities in the range $\omega_r/ \omega_{pc} \sim 0.2 - 0.4$, our minimum correlation interval is in the range $\omega_{pc} \tau_{min} \sim 15-30$.   FPC is evaluated at $x/\lambda_{Dc} = 0$. In the fiducial run, a pronounced oscillatory energy exchange occurs at short correlation times ($\tau$), but no single $\tau$ value entirely removes the oscillatory energy transfer for electrons. However, evidence of secular energy transfer is observed for both the beam (energy decrease) and the core (energy increase) protons.
The FPC map in Fig. \ref{fig:fiducial_phase}, panel a and b, reveals that both core and beam exchange energy with the fields in an oscillatory pattern over a large range of velocities: at fixed velocity, positive and negative signatures follow each other in time. Secular energy exchange is observed at the resonant velocity (dashed vertical line, refer to Table~\ref{comparison} for resonant velocities in all runs): the beam loses energy (negative blue signature, positive $\partial f_{pj} / \partial v$ in Eq. ~\ref{eq:FPC}), which drives the instability. Instead, the core gains energy (positive red signature, negative positive $\partial f_{pj} / \partial v$ in Eq. ~\ref{eq:FPC}). We do not observe secular energy exchange for the electrons, panel c, which just exhibit an oscillator pattern. In panel d we integrate energy exchange for the different particle populations in velocity, at the same spatial position, $x / \lambda_{Dc}=0$ and using $\tau \omega_{pc}= 100$: we observe an oscillatory behavior for beam protons and electrons, but not for the core. We will observe in subsequent analysis (Figs.~\ref{fpc1},~\ref{fpc2}, ~\ref{fpc3} and~\ref{fig:PosVar}) that oscillations are visible in the energy evolution of electrons and beam when the growth rate is low.

\subsection{Series One: Varying Electron Temperature}
\label{ssec:results.run1}
Figs.\ref{netenergy1} and~\ref{fpc1} refer to Series One simulations, where we decrease the electron-to-core temperature ratio. This quantity ($T_e/T_c$) significantly alters the IIAI growth rate and hence the energy exchange dynamics. This dependence arises from the shift of the ion-acoustic phase speed, and therefore of the resonance velocity, relative to the proton velocity distributions. While the pattern of energy exchange is consistent with the fiducial run (the beam secularly loses energy, the core gains it), the amount of exchanged energy decreases significantly with decreasing $T_e/T_c$ (row 1 to 3), and hence decreasing growth rate, as the resonance velocity moves toward regions of weaker positive velocity-space gradient in the beam distribution.
$\int dv FPC$ (Fig.~\ref{netenergy1}) shows that oscillatory energy exchange diminishes with longer $\tau$, shifting toward secular transfer: with longer correlation intervals, oscillations in energy transfer are averaged out. The FPC plots (Fig.~\ref{fpc1}) are consistent with those for the fiducial run, confirming that proton dynamics remain tied to ion-acoustic resonances occurring near the resonance velocity.
When the IIAI growth rate is higher for larger $T_e/T_c$, the resonance velocity lies in a region of strong positive slope of the beam distribution, resulting in efficient secular energy transfer. In this case, oscillatory energy exchange is not visible in the $v$-integrated energy exchange plots in Fig.~\ref{fpc1}, panel d. It becomes visible with decreasing growth rate, as the resonance velocity shifts away from this region and secular exchange weakens.

\subsection{Series Two: Varying Proton Beam Density}
\label{ssec:results.run2}
Fig.~\ref{netenergy2} and Fig.~\ref{fpc2} refer to Series 2 simulations, where we reduce the beam-to-core density ratio. We observe reduced energy exchange with decreasing beam density, and hence a reduced growth rate, as fewer beam particles are available to resonate with the ion-acoustic wave at the resonance velocity. As expected, secular energy exchange for both beam and core populations occurs at the resonant velocity. In all three cases explored here, the IIAI growth rate remains sufficiently high that the secular energy exchange dominates and masks the oscillatory energy exchange behavior in panels d of Fig.~\ref{fpc2}.

\subsection{Run Three: Varying Drift Speed}
\label{ssec:results.run3}

Figs.~\ref{netenergy3} and~\ref{fpc3} refer to Series 3, where we reduce the beam drift speed. The observed behavior is consistent with that shown in Figs.~\ref{netenergy2} and~\ref{netenergy3}, and in Figs.~\ref{fpc2} and~\ref{fpc3}. In this parameter scan, changes in the drift speed primarily modify the relative position of the resonance velocity with respect to the beam distribution, leading to variations in growth rate and energy exchange that follow the same resonance-controlled trends observed in the previous series.

\subsection{Energy Transfer as Functions of Simulation Position}

Until now, we have evaluated FPC at a fixed position, $x/\lambda_{Dc}=0$. Now we want to examine how FPC traces change in different areas of the domain. For this analysis, we use Run 2b, where the secular energy exchange between species is high enough to mask oscillatory energy exchange. Similar considerations hold for the other simulations. 
In Fig.~\ref{fig:PosVar} we depict the velocity-integrated FPC as a function of time 
at position $x_0/ \lambda_{Dc}= 0, 12.5, 25, 37.5$ respectively. We see that the core proton signature does not change significantly with position. The beam signatures stay negative after instability onset, with values at a fixed time changing slightly as a function of the sampling point position. The electron signature is the one that changes more significantly after instability onset, from positive at position $x_0/ \lambda_{Dc}= 0, 37.5$ to close to zero at $x_0/ \lambda_{Dc}= 12.5, 25$. We understand the reason for this behavior from Fig.~\ref{fig:PS2B}, where we plot the phase space for beam, core and electrons at time $\omega_{pc}t= 130$, marked as a horizontal line in Fig.~\ref{fig:PosVar}. Out of completeness, we plot in panel d the electric field as a function of space. The location of the sampling points in Fig.~\ref{fig:PosVar} are marked as vertical lines in Fig.~\ref{fig:PS2B}. We see that the core phase space changes minimally with $x$ in Fig.~\ref{fig:PS2B}, panel b: as already highlighted e.g. in~\citet{Afify2025}, the core is minimally affected by IIAI at least in these parameter regimes. In panel a, we see that the beam protons develop into an ion hole, which is obviously spatially dependent: this is the reason of the slightly variation of the beam signature with space in Fig.~\ref{fig:PosVar}. The electrons react to the ion hole and to the corresponding electric field by exhibiting slightly increase phase space density in correspondence with the electric field zero at $x/ \lambda_{Dc} \sim 7.8$. This spatial variation reflects in Fig.~\ref{fig:PosVar}. The dependence of FPC signatures with position, and their relation with nodes/antinodes of the fields in the case of instabilities has already been highlighted in \citep{Klein2017b}.

\begin{figure*}  [h!]
\centering
     \begin{minipage}[b]{1\textwidth}
    \includegraphics[width=\textwidth]{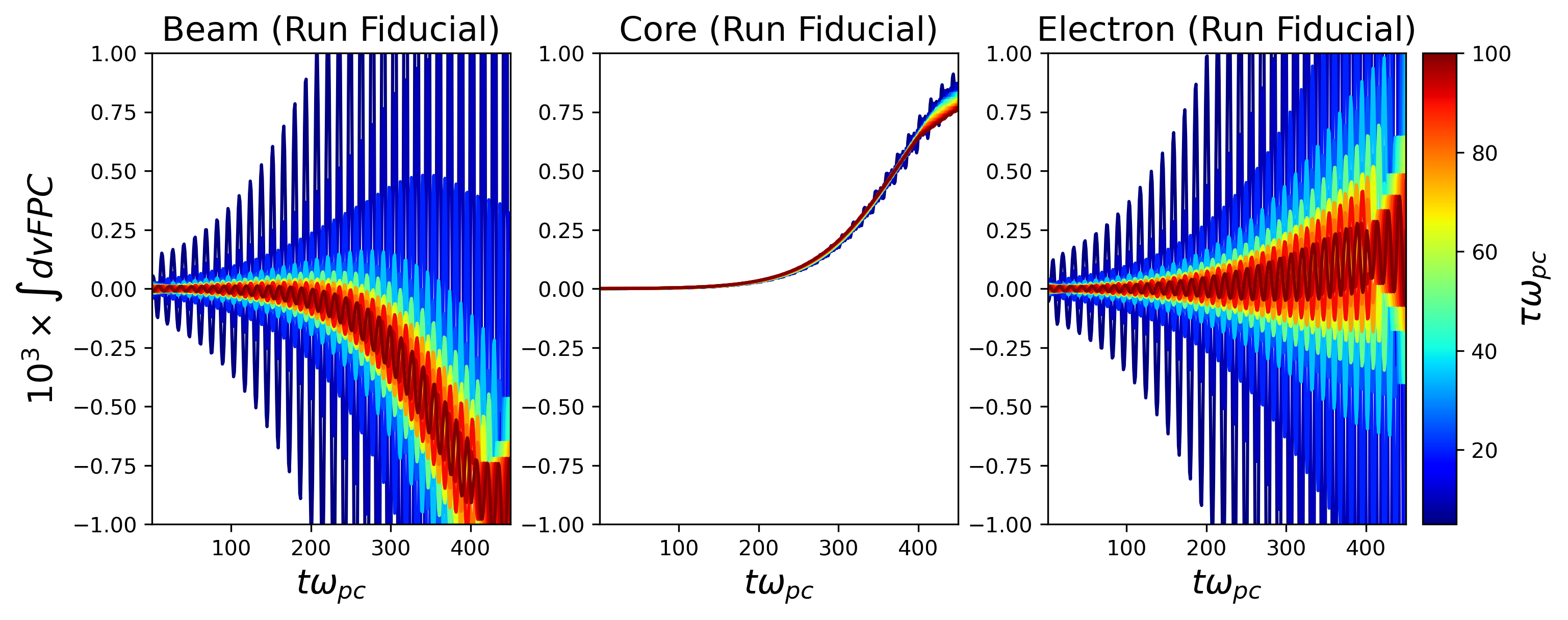}
  \end{minipage}
\caption{$\int dv\,FPC$ for a range of correlation intervals of RUN Fiducial at $x/\lambda_{Dc}=0$.} 
\label{fig:fiducial_fpc}
\end{figure*}

\begin{figure*}  [h!]
     \centering
     \begin{minipage}[b]{0.9\textwidth}
    \includegraphics[width=\textwidth]{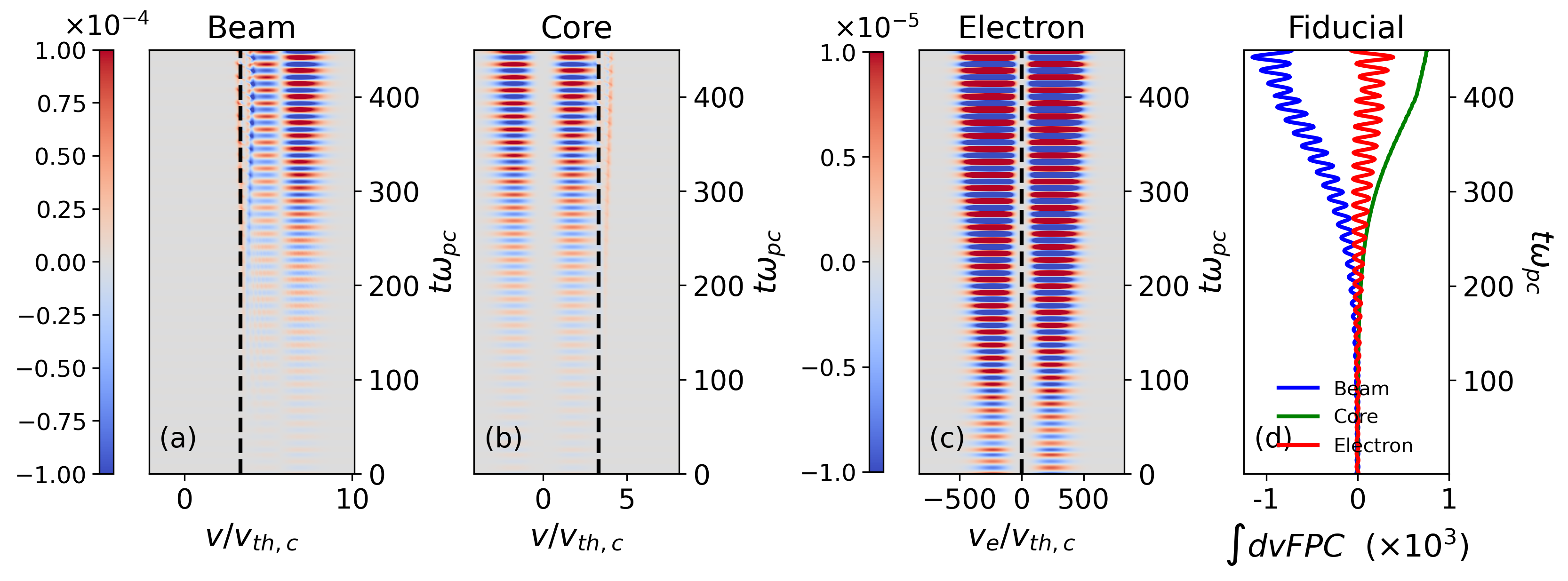}
  \end{minipage}
     \caption{The velocity-dependent FPC for RUN Fiducial at $x/\lambda_{Dc}=0$ for beam, core and electrons (panel a, b, c) for a correlation interval of $\tau \omega_{pc}=30$. Panel d: $\int dv\,FPC$ 
     The correlation interval $\tau \omega_{pc}$ is set to 100.
     The vertical dashed lines indicate the resonant velocity, which is 3.333 $v_{th,c}$.} 
    \label{fig:fiducial_phase}
\end{figure*}

\begin{figure*}  [h!]
\includegraphics[scale=0.37]{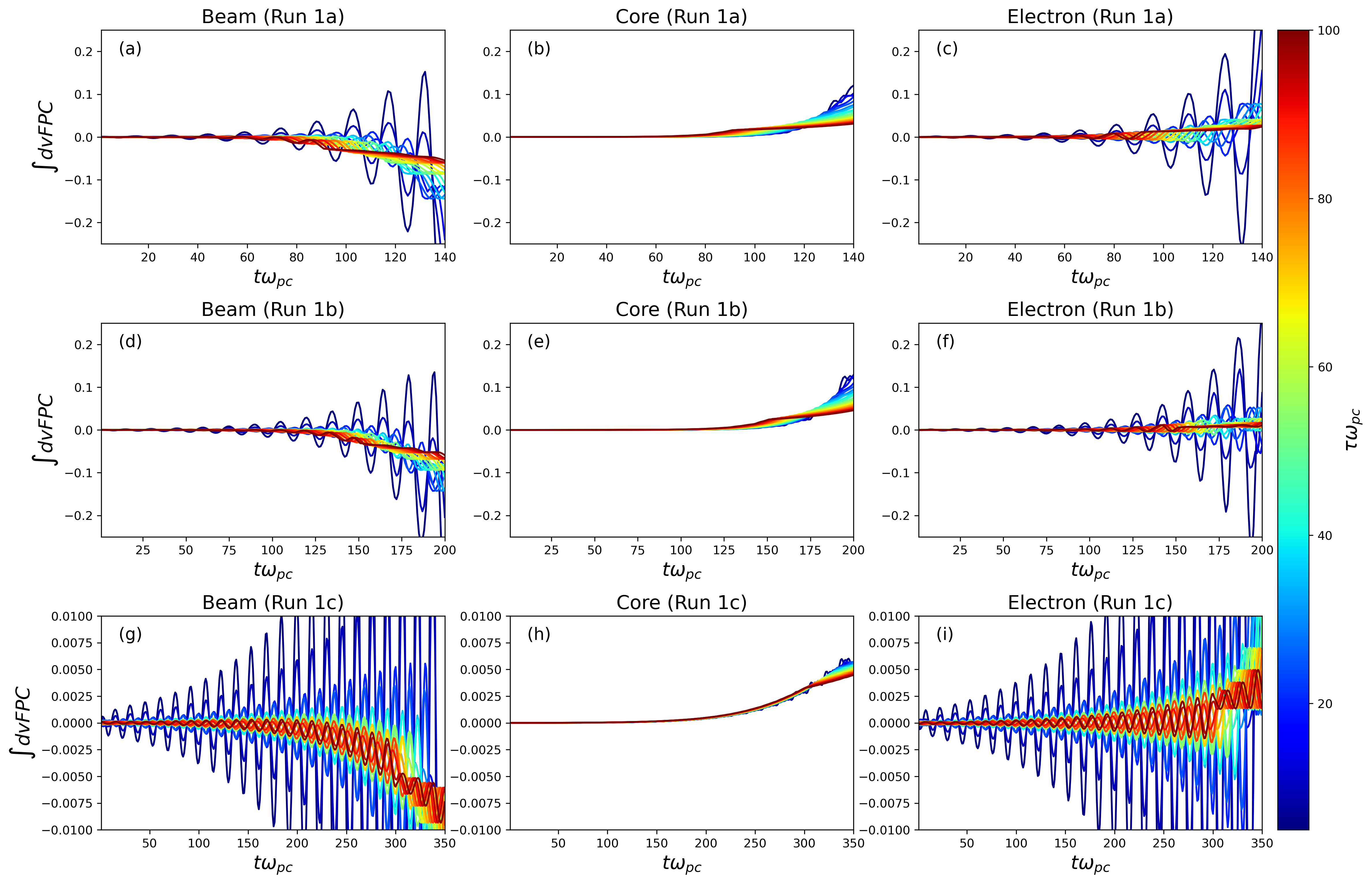}
\caption{$\int dv FPC$  for a range of correlation intervals for Series 1 tuns at $x/\lambda_{Dc}=0$. 
Run 1, 2, 3 are shown in three rows from top to bottom, respectively.} 
\label{netenergy1}
\end{figure*}

\begin{figure*}  [h!]
     \centering
     \begin{minipage}[b]{0.8\textwidth}
    \includegraphics[width=\textwidth]{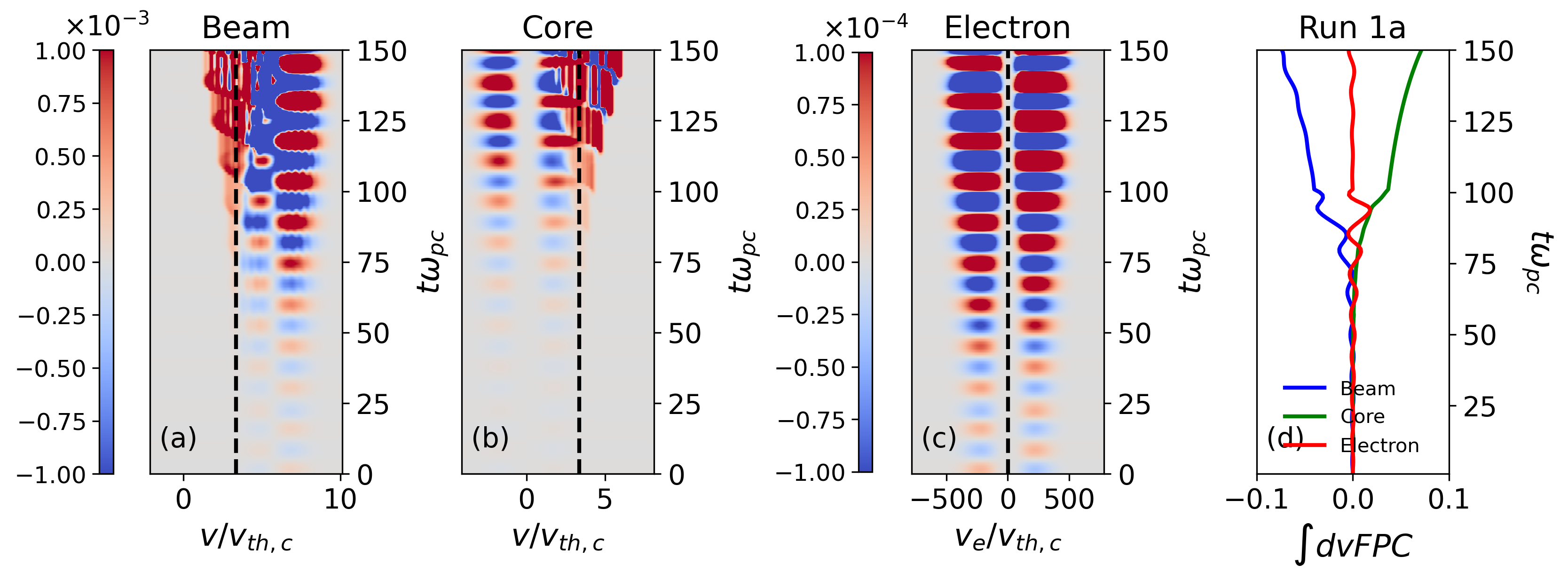}
  \end{minipage}
  \hfill
  \begin{minipage}[b]{0.8\textwidth}
    \includegraphics[width=\textwidth]{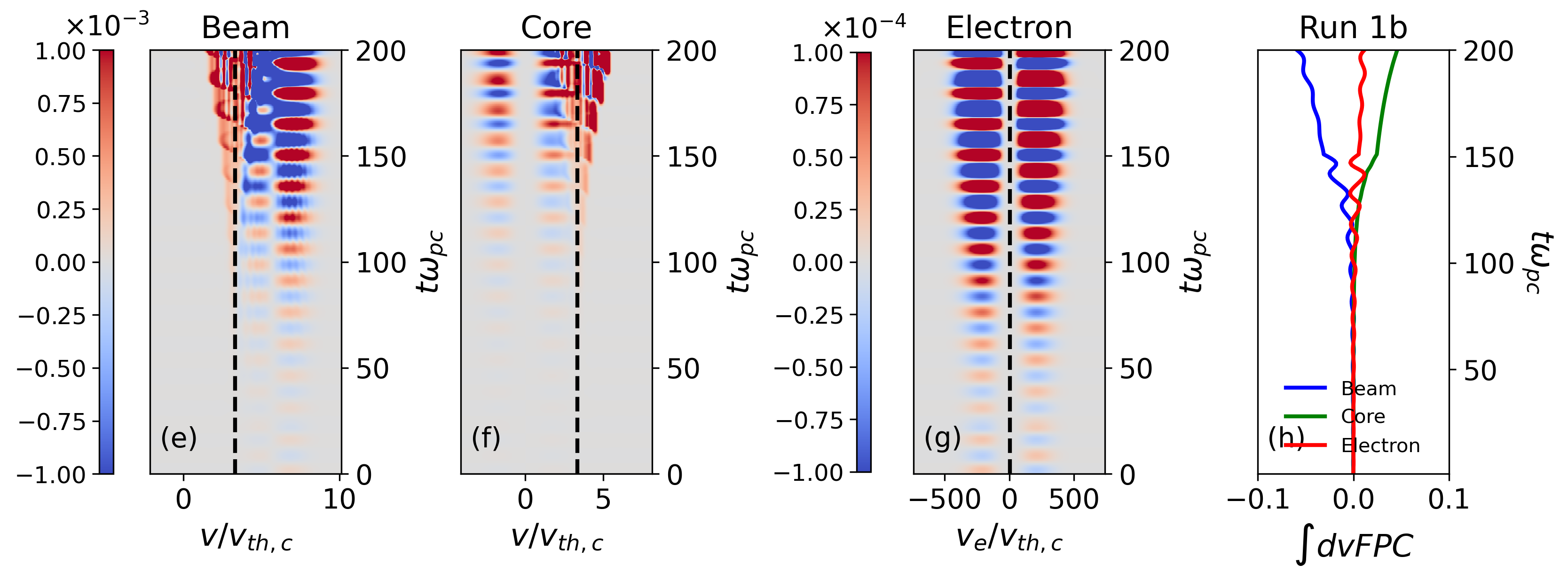}
  \end{minipage}
  \hfill
  \begin{minipage}[b]{0.8\textwidth}
    \includegraphics[width=\textwidth]{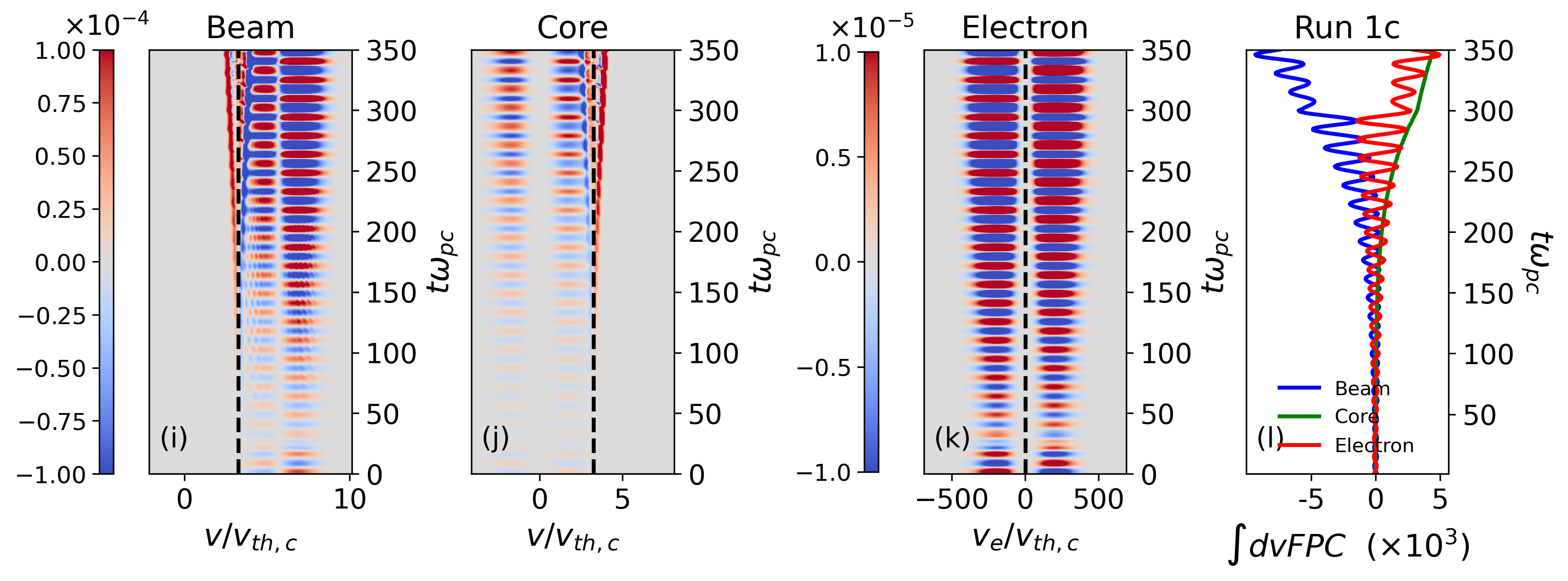}
  \end{minipage}
     \caption{The velocity-dependent FPC for Series 1 runs at $x/\lambda_{Dc}=0$. 
     The correlation interval $\tau \omega_{pc}$ is set to 100 for protons and 10 for electrons, except in Case 1c, where it is set to 10 for both protons and electrons. 
     The vertical dashed lines indicate the resonant velocities, which are 3.377, 3.333, and 3.284 from top to bottom, respectively.} 
     \label{fpc1}
\end{figure*}

\begin{figure*}  [h!]
\includegraphics[scale=0.37]{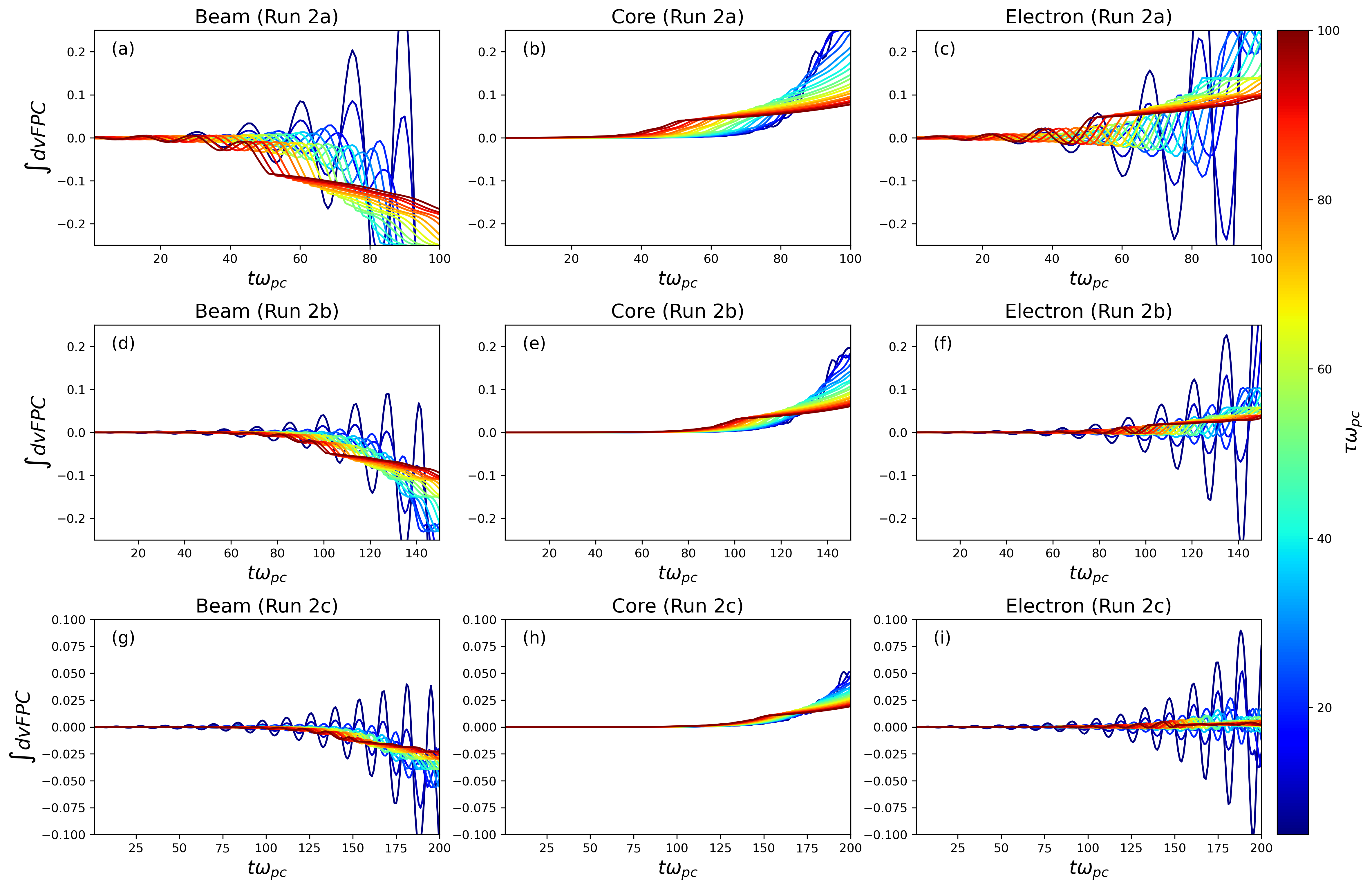}
\caption{The net energy transfer rate over a range of correlation intervals for Series 2 $x/\lambda_{Dc}=0$, arranged in the same format as in Fig. \ref{netenergy1}. } 
\label{netenergy2}
\end{figure*}

\begin{figure*}  [h!]
     \centering
     \begin{minipage}[b]{0.8\textwidth}
    \includegraphics[width=\textwidth]{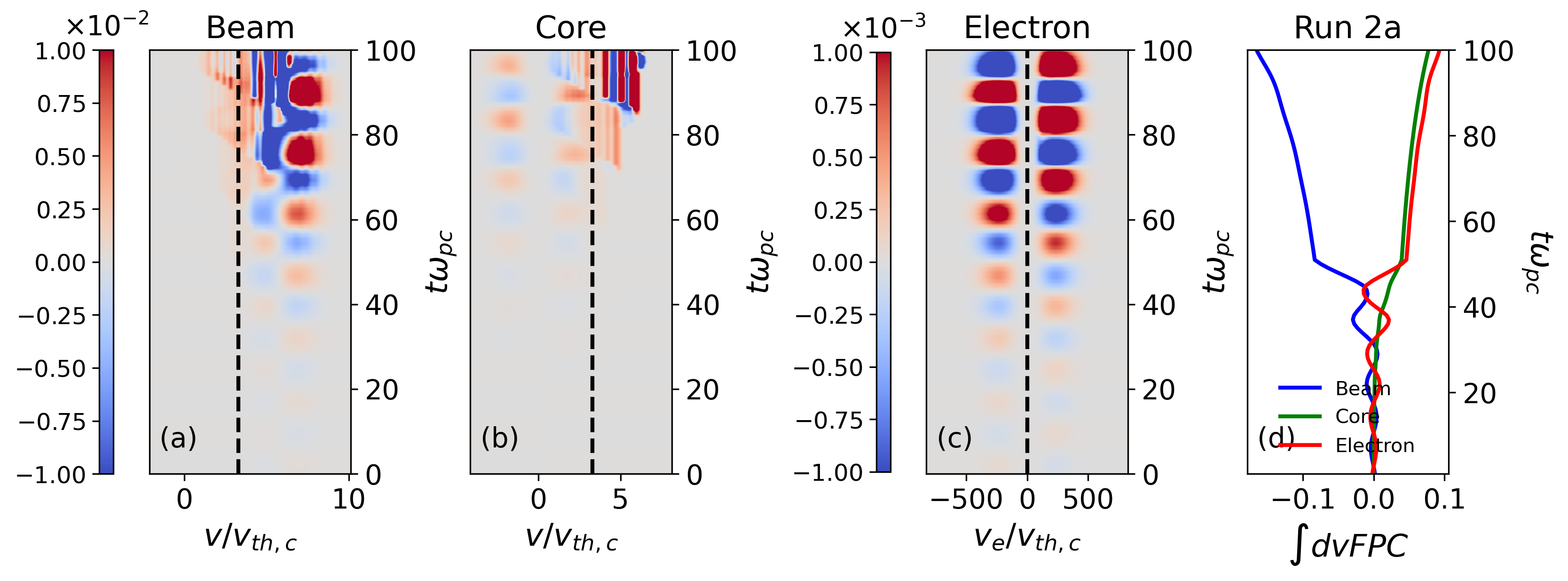}
  \end{minipage}
  \hfill
  \begin{minipage}[b]{0.8\textwidth}
    \includegraphics[width=\textwidth]{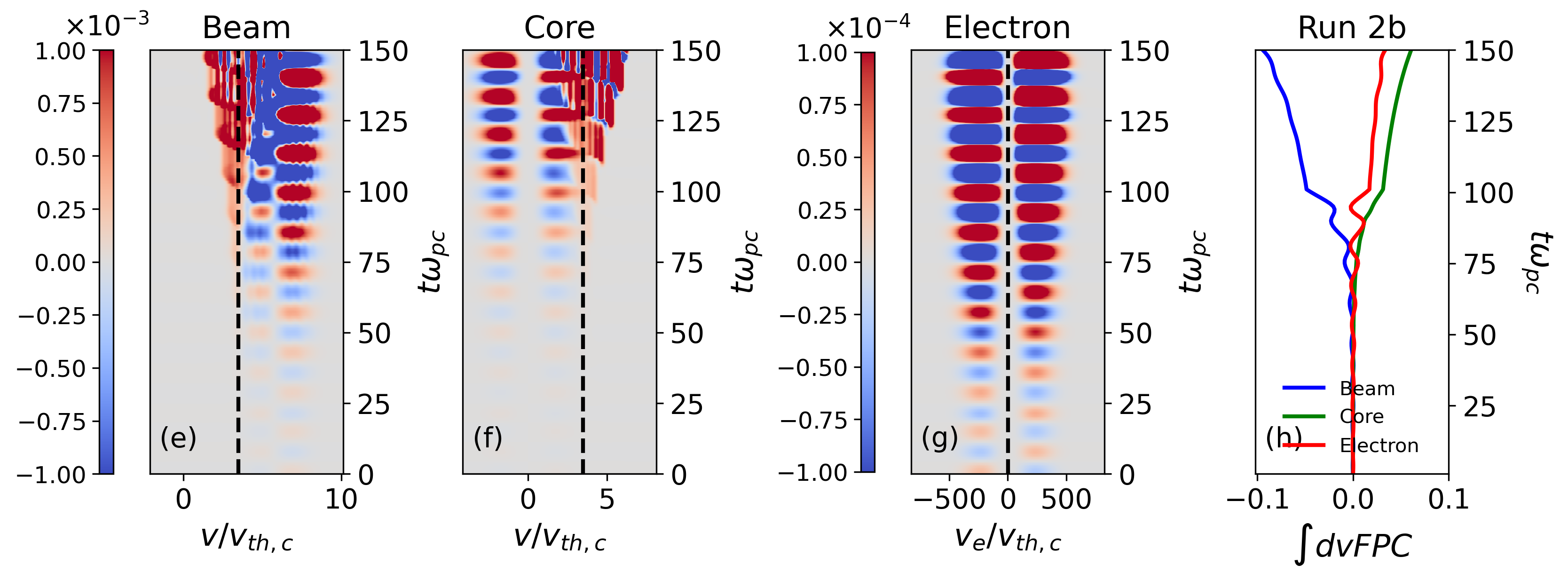}
  \end{minipage}
  \hfill
  \begin{minipage}[b]{0.8\textwidth}
    \includegraphics[width=\textwidth]{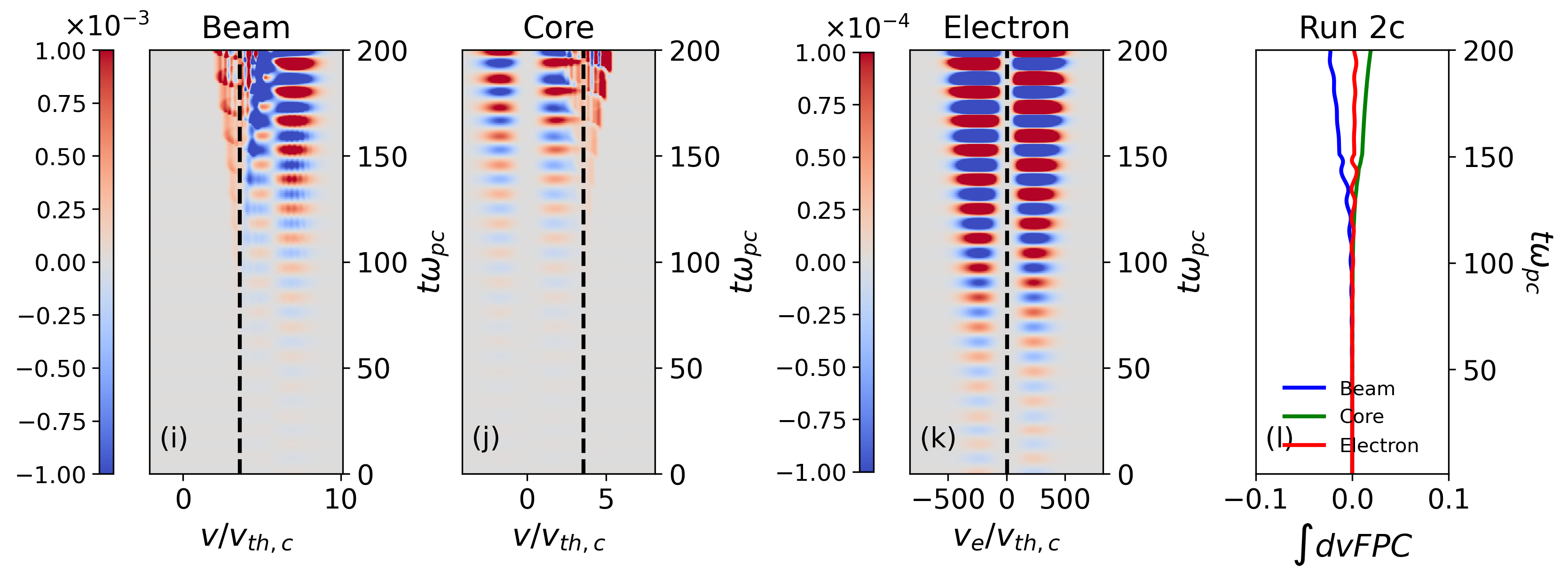}
  \end{minipage}
     \caption{The velocity-dependent FPC for Series 2 runs at $x/\lambda_{Dc}=0$. 
     The correlation interval $\tau \omega_{pc}$ is set to 100. 
     The vertical dashed lines indicate the resonant velocities, which are 3.295, 3.481, and 3.583 from top to bottom, respectively.} 
     \label{fpc2}
\end{figure*}

\begin{figure*}  [h!]
\includegraphics[scale=0.37]{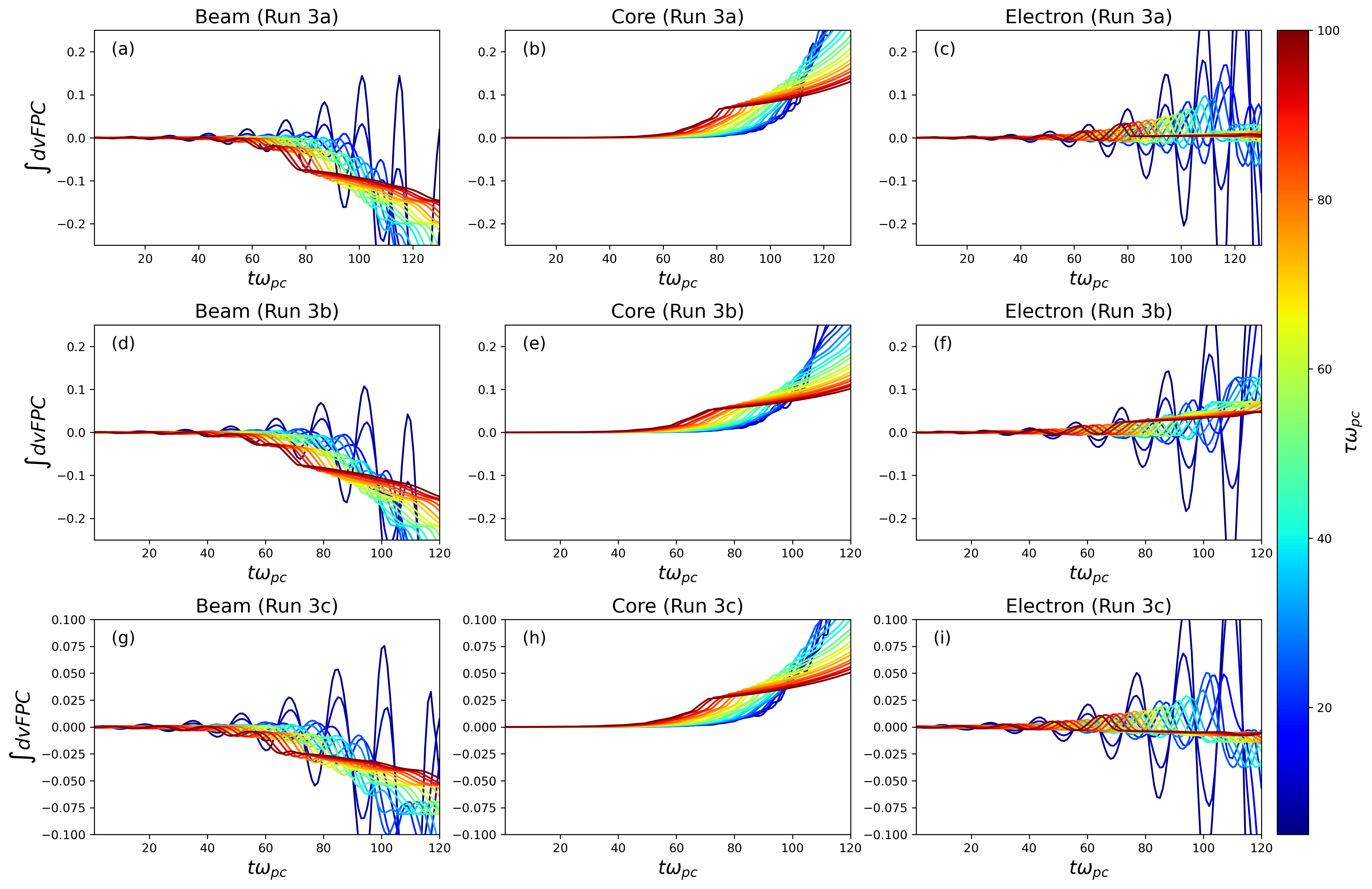}
\caption{$\int dv FPC$  for a range of correlation intervals of Series 3 runs at $x/\lambda_{Dc}=0$, arranged in the same format as in Fig. \ref{netenergy1}} 
\label{netenergy3}
\end{figure*}

\begin{figure*}  [h!]
     \centering
     \begin{minipage}[b]{0.8\textwidth}
    \includegraphics[width=\textwidth]{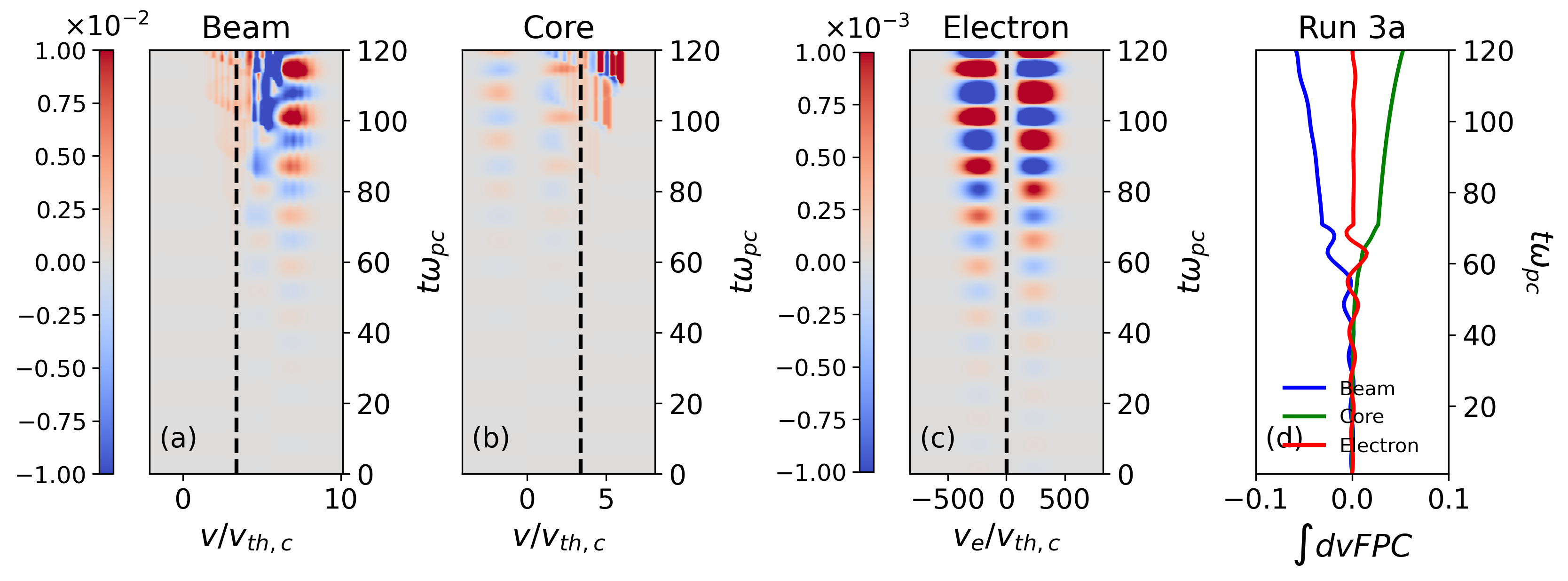}
  \end{minipage}
  \hfill
  \begin{minipage}[b]{0.8\textwidth}
    \includegraphics[width=\textwidth]{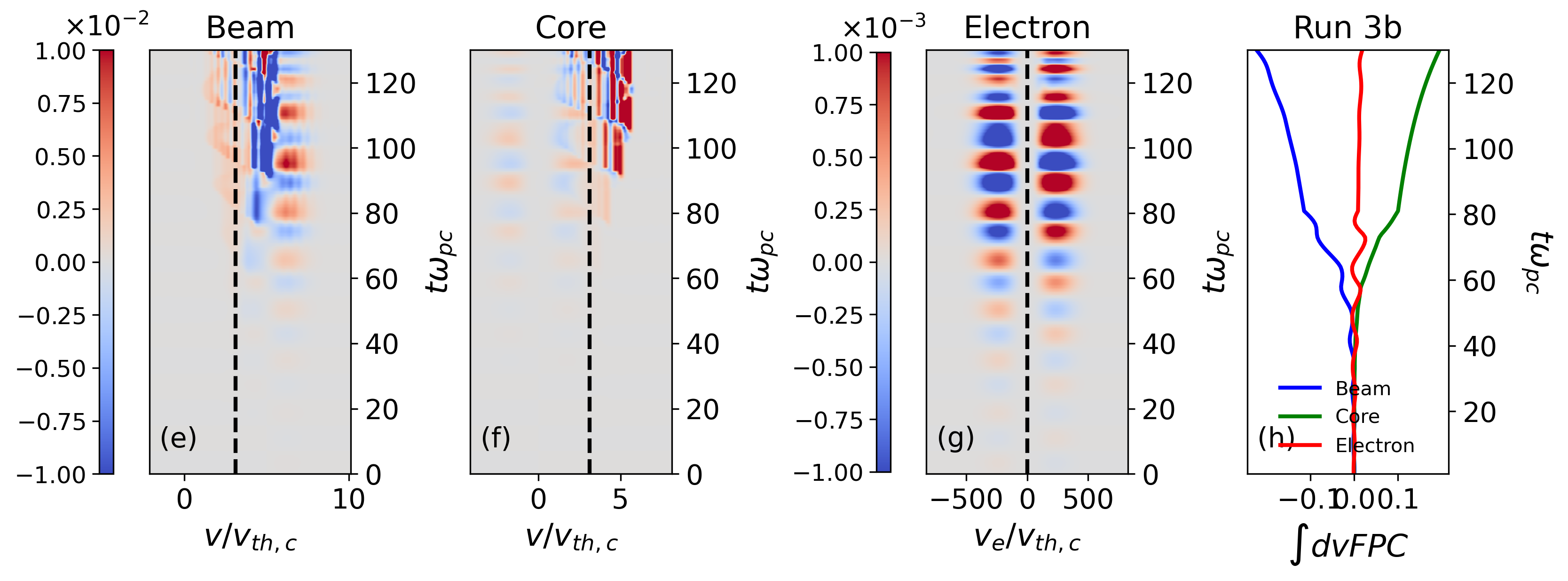}
  \end{minipage}
  \hfill
  \begin{minipage}[b]{0.8\textwidth}
    \includegraphics[width=\textwidth]{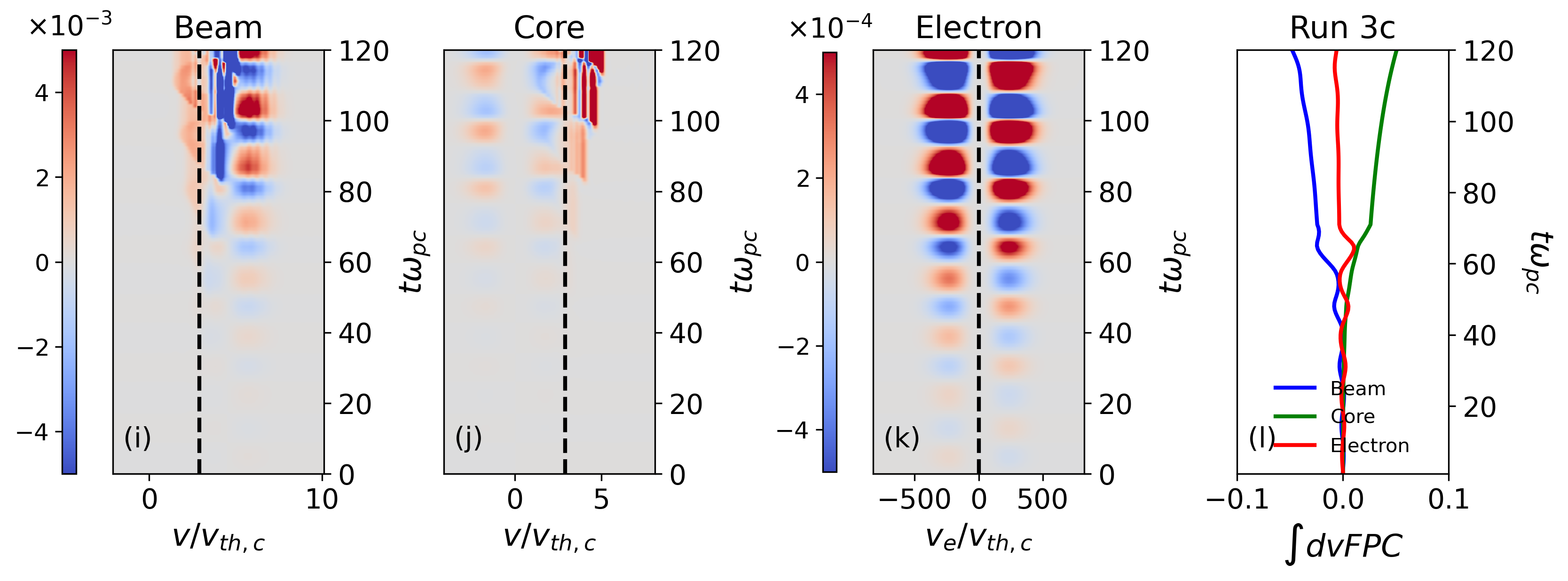}
  \end{minipage}
     \caption{The velocity-dependent FPC for Series 3 runs at $x/\lambda_{Dc}=0$. 
     The correlation interval $\tau \omega_{pc}$ is set to 100. 
     The vertical dashed lines indicate the resonant velocities, which are 3.408, 3.114, and 2.909 from top to bottom, respectively.} 
     \label{fpc3}
\end{figure*}

\section{Summary and Discussion}
\label{sec:discussion}

In this paper we applied FPC to fully kinetic simulations of the IIAI, often observed in the young solar wind. Our aim is to use fully kinetic simulations to identify characteristic IIAI FPC signatures, to make future identification of IIAI in observations more efficient. For this reason, our simulations are in parameter regimes compatible with observed solar wind properties. 
First, in Fig. \ref{fig:Eng_Fiducial} and \ref{fig:Eng_runs}, we examine energy transfer in the simulations, using both diagnostics readily available in simulations (Eq. \ref{normal}) and the non-linear wave particle interaction term which constitutes the base of FPC (Eq. \ref{barw}). Both approaches deliver the same results, since they are different ways of calculating the same quantity.
This analysis confirms that the proton beam is the primary instability driver, i.e. the particle population whose energy powers the instability (see Figs. \ref{fig:Eng_Fiducial} and \ref{fig:Eng_runs}). The majority of the energy lost by the beam population is gained by the core population, with energy transfer increasing with the growth rate of the instability (Fig. \ref{fig:Eng_runs}). The energy gained by the electron population is not negligible in absolute terms (Fig. \ref{fig:Eng_Fiducial}, upper panel), but small with respect to their initial energy (Fig. \ref{fig:Eng_Fiducial}, lower panel). 

\begin{figure*}[h!]
\centering
\begin{minipage}[b]{0.22\textwidth}
    \centering
    \includegraphics[width=\textwidth]{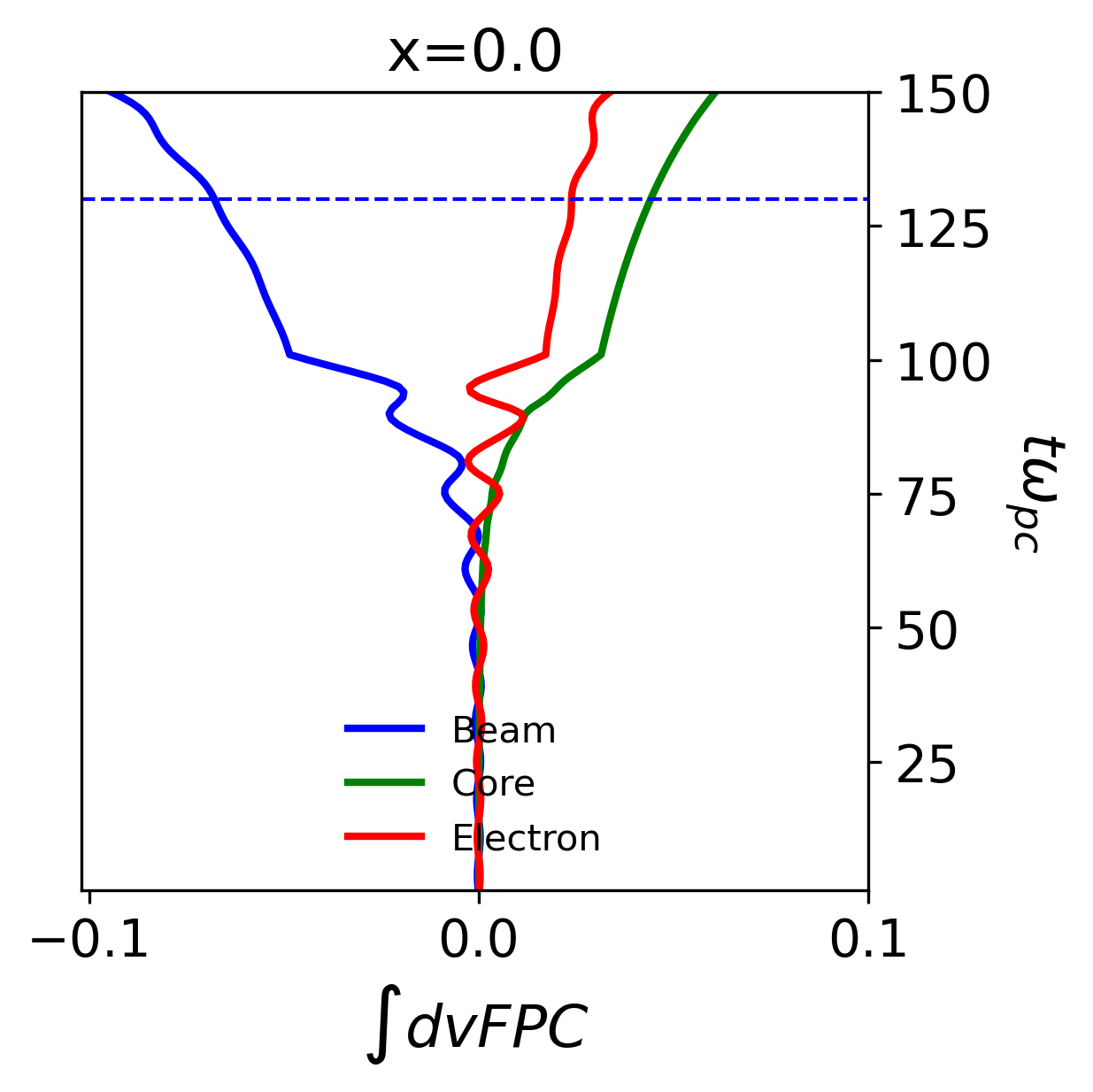}
\end{minipage}
\vspace{1em} 
\begin{minipage}[b]{0.22\textwidth}
    \centering
    \includegraphics[width=\textwidth]{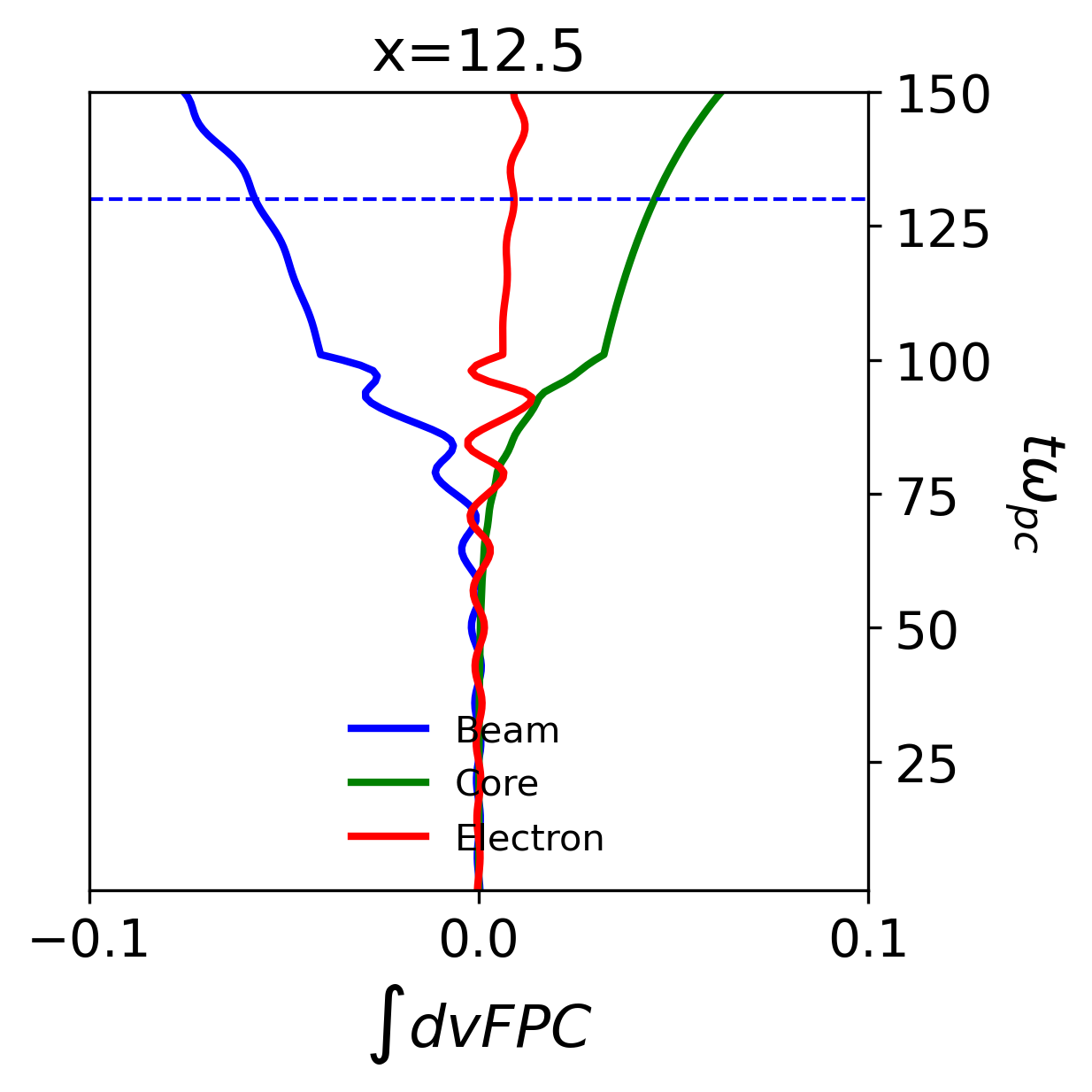}
\end{minipage}
\begin{minipage}[b]{0.22\textwidth}
    \includegraphics[width=\textwidth]{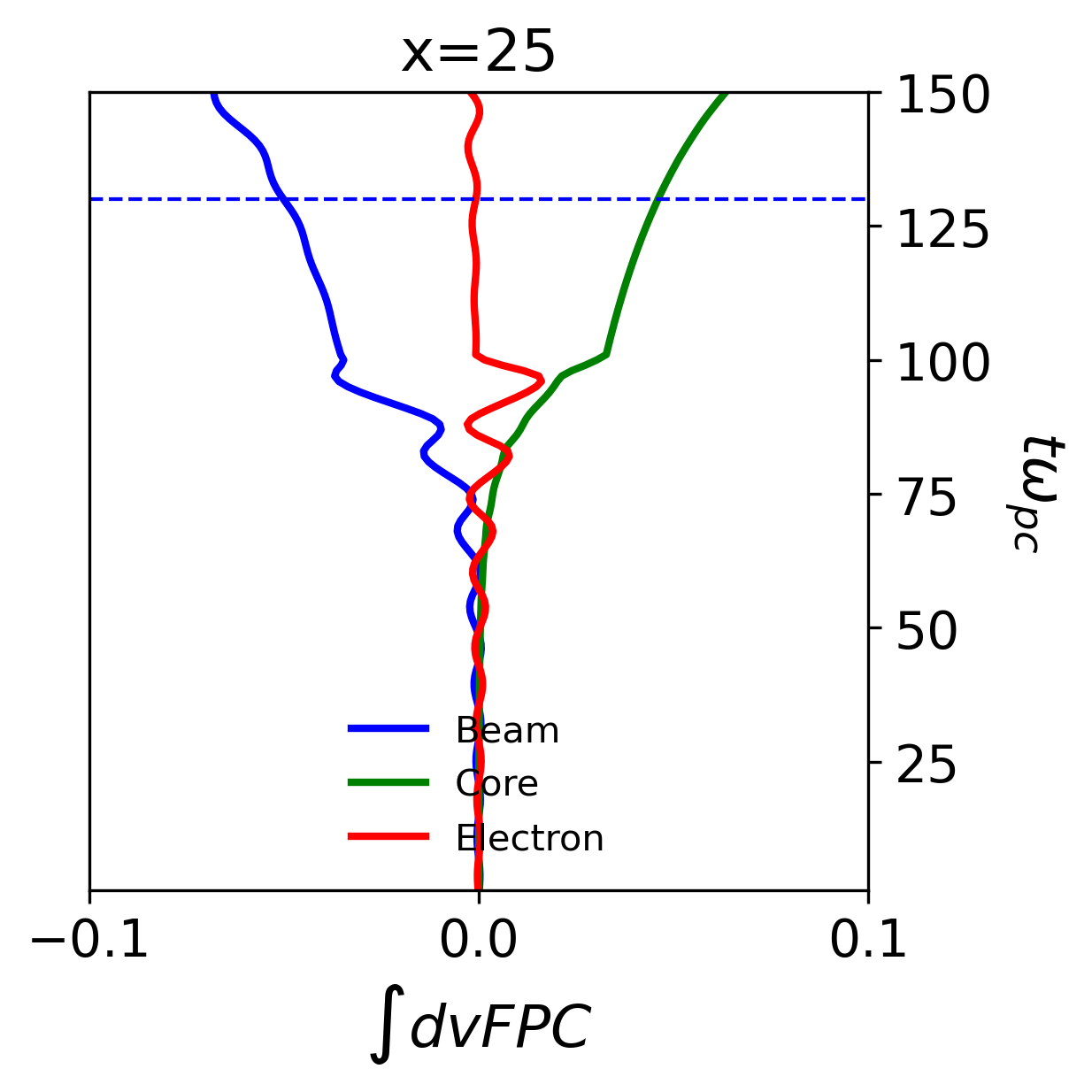}
\end{minipage}
\begin{minipage}[b]{0.22\textwidth}
    \includegraphics[width=\textwidth]{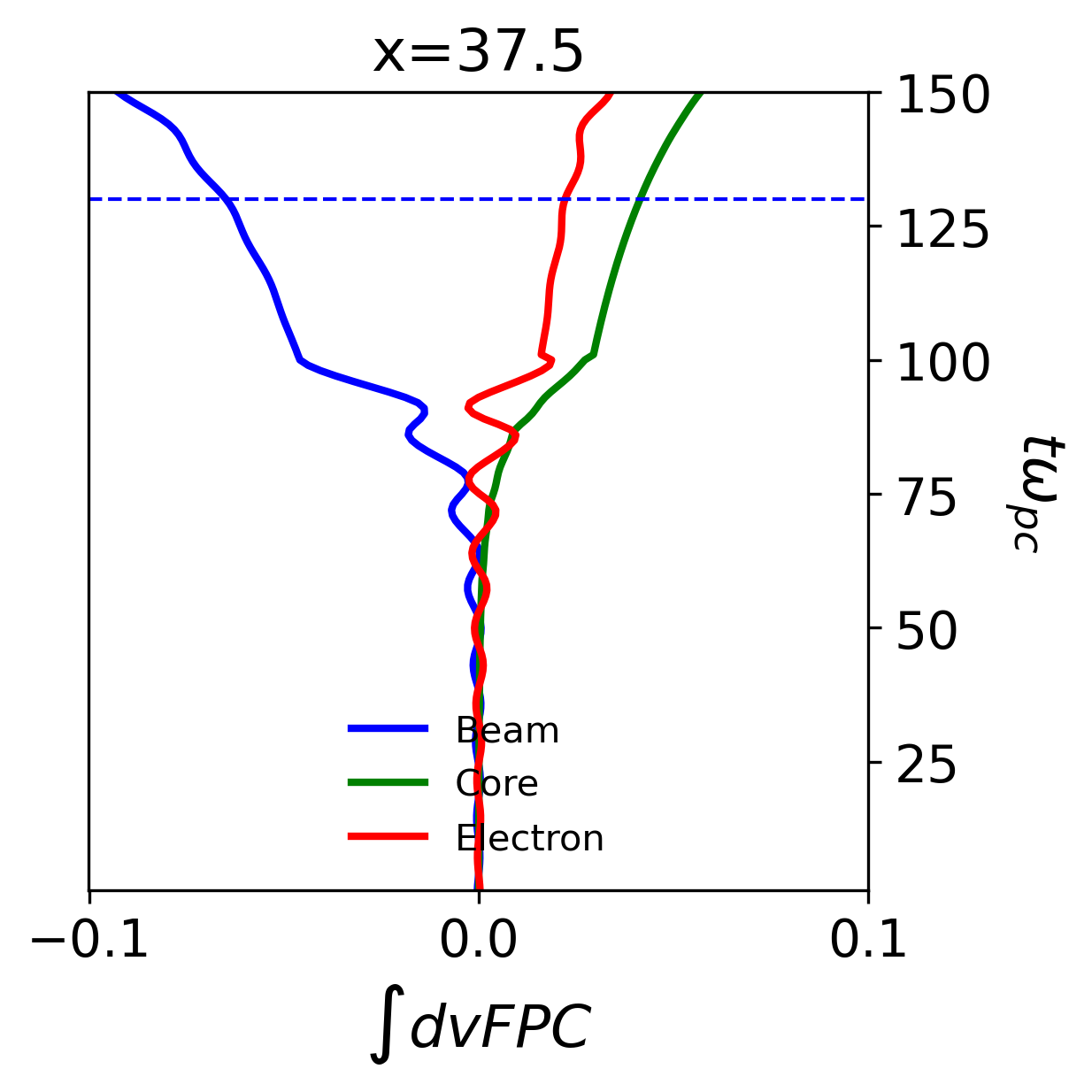}
\end{minipage}

\caption{Velocity integrated FPC as a function of time for run 2B, evaluated at position $x/ \lambda_{dc}=0, 12.5, 25, 37.5$ respectively. The horizontal line marks $\omega_{pc}t= 130.$}
\label{fig:PosVar}
\end{figure*}

\begin{figure}[h!]
\centering
\begin{minipage}[b]{0.85\textwidth}
    \centering
    \includegraphics[width=\textwidth]{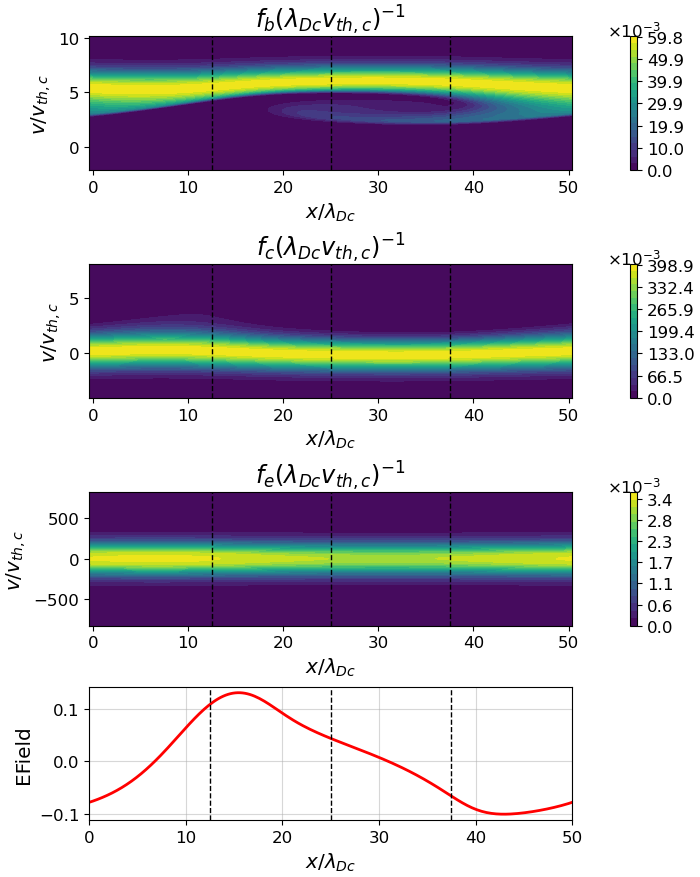}
\end{minipage}
\caption{Phase space for beam, core, electrons (panel a to c) and electric field (panel d) for the 2b run at $\omega_{pc}t= 130$. The vertical lines mark $x/ \lambda_{Dc}=0, 12.5, 25, 37.5$.}
\label{fig:PS2B}
\end{figure}

We examine the effects of using different FPC correlation intervals for our analysis (Figs. \ref{fig:fiducial_fpc}, \ref{netenergy1}, \ref{netenergy2}, \ref{netenergy3}). We find that $\tau=100\omega_{pc}$, which corresponds to $\tau/ T \sim 8$, with $T= 2 \pi / \omega$, is a value capable of highlighting secular vs oscillatory energy transfer.
We then look for characteristic FPC signatures for the different particle populations (Fig. \ref{fig:fiducial_phase}, \ref{fpc1}, \ref{fpc2}, \ref{fpc3}). Integrating FPC in velocity space (fourth panel in each row in Fig. \ref{fig:fiducial_phase}, \ref{fpc1}, \ref{fpc2}, \ref{fpc3}), we observe that the oscillations observed in the proton beam and electron signatures and associated with oscillatory energy exchange patterns are dominated by secular exchange when the IIAI growth rate is high. Examining FPC signatures as a function of time in velocity space (first three panels in Fig. \ref{fig:fiducial_phase}, \ref{fpc1}, \ref{fpc2}, \ref{fpc3}), we observe traces of secular and resonant energy transfer in correspondence of the IA resonance velocity for both the proton beam (first panel) and core (second panel): the beam loses energy, while the core gains it. The electrons (third panel), instead, exhibit only an oscillatory energy exchange pattern between the fields and the particles. Changing the key parameters that determine the growth rate ($T_e/Tc$, $n_b/n_c$, and $V_{D,b}/v_{th,c}$) we observe strong positive correlation between the growth rate and the the exchanged energy. The ratio of energy exchange between different particle components appear consistent across growth rates.  
Investigating the impact of probe location in simulations gives quite interesting results (Figs. \ref{fig:PosVar} and \ref{fig:PS2B}). We observe that energy transfer for the proton core population is not affected by probe location, since the core proton population is minimally affected by the formation of ion holes generated by the IIAI. Proton beam and electron results, instead, show position dependency, since both the proton beam and electron velocity distribution are affected by ion holes: ion holes are constituted of beam protons but modify the electron velocity distribution as well through electric field modulation (Fig. \ref{fig:PS2B}). 

Although the present analysis is not designed to model spacecraft sampling effects such as Doppler shifting and sampling frequencies, the results are nonetheless relevant for the interpretation of single-point measurements. A detailed quantification of these effects applied specifically to PSP and Solar Orbiter measurements will be addressed in future studies. The FPC technique is intrinsically a single-spacecraft diagnostic, depending only on local time series of the electric field and particle velocity distribution functions. The velocity-space signatures of IIAI identified here therefore provide a physically grounded reference for what such instabilities would produce in idealized single-point measurements.

A natural future outlook for this investigation is to look for the IIAI FPC signatures identified here in observations of the young solar wind. This task will be left for future missions supporting higher instrumental cadence, since Parker Solar Probe's SPANi fastest possible cadence is $0.87\,\mathrm{s}$ \citep{livi2022solar}. As remarked in \citet{malaspina2024frequency}, this is too slow to capture proton signatures of ongoing IIAI dynamics if the time scale of the IIAI (inverse of the growth rate) is of the order of tens of milliseconds, as obtained in this and previous work~\citep{Afify2024} in parameter ranges compatible with observations. Also, evolution times of the order of a few percent of the ion gyrofrequency are too fast for direct PSP sampling, if the gyrofrequency is calculated with magnetic field values of the order of $B= 200\;nT$, as reported in correspondence with IIAI observations in~\citep{Mozer2023}, Table 1. 

Future work will examine more realistic simulation scenarios that take into account both the anisotropies in the different particle populations frequently observed in the solar wind and the possible role of larger-scale processes, e.g. magnetic reconnection, in the production of ion beams that result into IIAI.

\section*{Acknowledgement}
M. S. Afify thanks the Alexander-von-Humboldt Foundation, 53173 Bonn, Germany (Ref 3.4-1229224-EGY-HFST-P) for the research fellowship and its financial support. M.E.I. acknowledges support from the Deutsche Forschungsgemeinschaft (DFG, German Research Foundation) within the Collaborative Research Center SFB1491 and project 497938371. We thank Jürgen Dreher for the useful discussions on the optimization of the Vlasov code used in this study.

\clearpage
\bibliographystyle{aasjournal}
\bibliography{main.bib}

\begin{thebibliography}{}
\expandafter\ifx\csname natexlab\endcsname\relax\def\natexlab#1{#1}\fi
\providecommand{\url}[1]{\href{#1}{#1}}
\providecommand{\dodoi}[1]{doi:~\href{http://doi.org/#1}{\nolinkurl{#1}}}
\providecommand{\doeprint}[1]{\href{http://ascl.net/#1}{\nolinkurl{http://ascl.net/#1}}}
\providecommand{\doarXiv}[1]{\href{https://arxiv.org/abs/#1}{\nolinkurl{https://arxiv.org/abs/#1}}}

\bibitem[{Afify {et~al.}(2025)Afify, Dreher, O'Neill, \& Innocenti}]{Afify2025}
Afify, M.~S., Dreher, J., O'Neill, S., \& Innocenti, M.~E. 2025,
  \href{https://doi.org/10.1051/0004-6361/202556866}{A\&A}

\bibitem[{Afify {et~al.}(2024)Afify, Dreher, Schoeffler, Micera, \&
  Innocenti}]{Afify2024}
Afify, M.~S., Dreher, J., Schoeffler, K., Micera, A., \& Innocenti, M.~E. 2024,
  \href{https://doi.org/10.3847/1538-4357/ad644c}{APJ, 971, 93}

\bibitem[{Afshari {et~al.}(2021)Afshari, Howes, Kletzing, Hartley, \&
  Boardsen}]{afshari2021importance}
Afshari, A., Howes, G., Kletzing, C., Hartley, D., \& Boardsen, S. 2021,
  Journal of Geophysical Research: Space Physics, 126, e2021JA029578

\bibitem[{Afshari {et~al.}(2024)Afshari, Howes, Shuster, McGinnis, Martinovic,
  Klein, Kletzing, Hartley, \& Boardsen}]{Afshari2024}
Afshari, A., Howes, G., Shuster, J., {et~al.} 2024,
  \href{https://doi.org/10.1038/s41467-024-52125-8}{Nat. Commun., 15, 7870}

\bibitem[{Baumjohann \& Treumann(2012)}]{Baumjohann2012}
Baumjohann, W., \& Treumann, R.~A. 2012,
  \href{https://doi.org/10.1142/p015}{(World Scientific)}

\bibitem[{Bold{\'u} {et~al.}(2024)Bold{\'u}, Graham, Morooka, Andre,
  Khotyaintsev, Dimmock, P{\'\i}{\v{s}}a, Sou{\v{c}}ek, Maksimovic, Louarn,
  {et~al.}}]{Bolduu2024}
Bold{\'u}, J.~J., Graham, D., Morooka, M., {et~al.} 2024,
  \href{https://doi.org/10.1029/2024GL109956}{Geophys. Res. Lett., 51,
  e2024GL109956.}

\bibitem[{Brown {et~al.}(2023)Brown, Juno, Howes, Haggerty, \&
  Constantinou}]{brown2023isolation}
Brown, C.~R., Juno, J., Howes, G.~G., Haggerty, C.~C., \& Constantinou, S.
  2023, Journal of Plasma Physics, 89, 905890308

\bibitem[{Cattell {et~al.}(2022)Cattell, Breneman, Dombeck, Hanson, Johnson, \&
  Halekas}]{Cattell2022}
Cattell, C., Breneman, A., Dombeck, J., {et~al.} 2022,
  \href{https://doi.org/10.3847/2041-8213/ac4015}{ApJL, 924, L33.}

\bibitem[{Chen {et~al.}(2019)Chen, Klein, \& Howes}]{Chen2019}
Chen, C., Klein, K., \& Howes, G.~G. 2019,
  \href{https://doi.org/10.1038/s41467-019-08435-3}{Nat. Commun., 10, 740}

\bibitem[{Dai {et~al.}(2021)Dai, Wang, \& Lavraud}]{Dai2021}
Dai, L., Wang, C., \& Lavraud, B. 2021,
  \href{https://doi.org/10.3847/1538-4357/ac0fde}{ApJ, 919, 15}

\bibitem[{Fox {et~al.}(2016)Fox, Velli, Bale, Decker, Driesman, Howard, \&
  Kasper}]{Fox2016}
Fox, N.~J., Velli, M.~C., Bale, S.~D., {et~al.} 2016,
  \href{https://doi.org/10.1007/s11214-015-0211-6}{SSRv, 204, 7}

\bibitem[{Fried \& Conte(1961)}]{fried1961plasma}
Fried, B.~D., \& Conte, S.~D. 1961,
  \href{https://doi.org/10.1016/C2013-0-12176-9}{The plasma dispersion
  function: the Hilbert transform of the Gaussian (Elsevier Inc)}

\bibitem[{Gary(1993)}]{GaryBook}
Gary, S.~P. 1993,
  \href{https://www.cambridge.org/us/catalogue/catalogue.asp?isbn=0521431670}{Theory
  of Space Plasma Microinstabilities (New York: Cambridge Univ. Press)}

\bibitem[{Gary \& Omidi(1987)}]{Gary1987}
Gary, S.~P., \& Omidi, N. 1987,
  \href{https://doi.org/10.1017/S0022377800011983}{J. Plasma Physics, 37,
  45-61.}

\bibitem[{Gonz{\'a}lez {et~al.}(2023)Gonz{\'a}lez, Innocenti, \&
  Tenerani}]{gonzalez2023particle}
Gonz{\'a}lez, C., Innocenti, M.~E., \& Tenerani, A. 2023,
  \href{https://doi.org/10.1017/S0022377823000120}{J. Plasma Phys., 89,
  905890208}

\bibitem[{Goodrich {et~al.}(2019)Goodrich, Ergun, Schwartz, Wilson~III,
  Johlander, Newman, Wilder, Holmes, Burch, Torbert,
  {et~al.}}]{goodrich2019impulsively}
Goodrich, K.~A., Ergun, R., Schwartz, S.~J., {et~al.} 2019,
  \href{https://doi.org/10.1029/2018JA026436}{J. Geophys. Res. Space Phys. 124,
  1855-1865.}

\bibitem[{Graham {et~al.}(2025{\natexlab{a}})Graham, Cozzani, Khotyaintsev,
  Wilder, Holmes, Nakamura, B{\"u}chner, Dokgo, Richard, Steinvall,
  {et~al.}}]{Graham2025b}
Graham, D., Cozzani, G., Khotyaintsev, Y.~V., {et~al.} 2025{\natexlab{a}},
  \href{https://doi.org/10.1007/s11214-024-01133-7}{Space Sci. Rev., 221, 20.}

\bibitem[{Graham {et~al.}(2025{\natexlab{b}})Graham, Khotyaintsev, \&
  Lalti}]{Graham2025IAWs}
Graham, D.~B., Khotyaintsev, Y.~V., \& Lalti, A. 2025{\natexlab{b}},
  \href{https://doi.org/10.48550/arXiv.2502.07953}{arXiv preprint
  arXiv:2502.07953}

\bibitem[{Graham {et~al.}(2021)Graham, Khotyaintsev, Sorriso-Valvo, Maksimovic,
  Vaivads, Soucek, Edberg, \& Píša}]{Graham2021}
Graham, D.~B., Khotyaintsev, Y.~V., Sorriso-Valvo, L., {et~al.} 2021,
  \href{https://doi.org/10.1051/0004-6361/202140943}{A\&A, 656, A23}

\bibitem[{Gurnett(1991)}]{Gurnett1991}
Gurnett, D.~A. 1991,
  \href{https://link.springer.com/chapter/10.1007/978-3-642-75364-0_4}{in Waves
  and Instabilities, eds. R. Schwenn, \& E. Marsch (Berlin, Heidelberg:
  Springer-Verlag), 135}

\bibitem[{Gurnett \& Anderson(1977)}]{Gurnett1977}
Gurnett, D.~A., \& Anderson, R.~R. 1977,
  \href{https://doi.org/10.1029/JA082i004p00632}{J. Geophys. Res. 82,
  632–650}

\bibitem[{Hollweg(1975)}]{hollweg1975waves}
Hollweg, J.~V. 1975, \href{https://doi.org/10.1029/RG013i001p00263}{Rev
  Geophys, 13, 263-289}

\bibitem[{Howes(2024)}]{howes2024fundamental}
Howes, G.~G. 2024, \href{https://doi.org/10.1017/S0022377824001090}{J. Plasma
  Phys., 90, 905900504.}

\bibitem[{Howes {et~al.}(2025)Howes, Felix, Brown, Haggerty, Juno, TenBarge,
  Wilson, \& Caprioli}]{howes2025velocity}
Howes, G.~G., Felix, A., Brown, C.~R., {et~al.} 2025,
  \href{https://doi.org/10.1063/5.0269528}{Phys. Plasmas, 32, 6.}

\bibitem[{Howes {et~al.}(2017)Howes, Klein, \& Li}]{howes2017diagnosing}
Howes, G.~G., Klein, K.~G., \& Li, T.~C. 2017,
  \href{https://doi.org/10.1017/S0022377816001197}{J. Plasma Phys., 83,
  705830102.}

\bibitem[{Huang {et~al.}(2024)Huang, Howes, \& McCubbin}]{huang2024velocity}
Huang, R., Howes, G.~G., \& McCubbin, A.~J. 2024,
  \href{https://doi.org/10.1017/S0022377824000667}{J. Plasma Phys., 90,
  535900401.}

\bibitem[{Juno {et~al.}(2023)Juno, Brown, Howes, Haggerty, TenBarge,
  Wilson~Iii, Caprioli, \& Klein}]{juno2023phase}
Juno, J., Brown, C.~R., Howes, G.~G., {et~al.} 2023, The Astrophysical Journal,
  944, 15

\bibitem[{Klein(2017)}]{Klein2017b}
Klein, K.~G. 2017, \href{https://doi.org/10.1063/1.4977465}{Phys. Plasmas, 24,
  5}

\bibitem[{Klein \& Howes(2016)}]{Klein2016}
Klein, K.~G., \& Howes, G.~G. 2016,
  \href{http://dx.doi.org/10.3847/2041-8205/826/2/L30}{APJL, 826, L30}

\bibitem[{Klein {et~al.}(2017)Klein, Howes, \& TenBarge}]{Klein2017a}
Klein, K.~G., Howes, G.~G., \& TenBarge, J.~M. 2017,
  \href{https://doi.org/10.1017/S0022377817000563}{JPlPh, 83, 535830401}

\bibitem[{Klein {et~al.}(2020)Klein, Howes, TenBarge, \&
  Valentini}]{klein2020diagnosing}
Klein, K.~G., Howes, G.~G., TenBarge, J.~M., \& Valentini, F. 2020,
  \href{https://doi.org/10.1017/S0022377820000689}{J. Plasma Phys., 86,
  905860402.}

\bibitem[{Kurth {et~al.}(1979)Kurth, Gurnett, \& Scarf}]{Kurth1979}
Kurth, W., Gurnett, D., \& Scarf, F. 1979,
  \href{https://doi.org/10.1029/JA084iA07p03413}{J. Geophys. Res. 82,
  632–650}

\bibitem[{Li {et~al.}(2025)Li, Liu, \& Loureiro}]{li2025role}
Li, D., Liu, Z., \& Loureiro, N.~F. 2025,
  \href{https://doi.org/10.48550/arXiv.2505.08983}{arXiv preprint
  arXiv:2505.08983}

\bibitem[{Liu {et~al.}(2024)Liu, White, Francisquez, Milanese, \&
  Loureiro}]{Liu2024a}
Liu, Z., White, R., Francisquez, M., Milanese, L.~M., \& Loureiro, N.~F. 2024,
  \href{https://doi.org/10.1017/S0022377824000060}{J. Plasma Phys., 90,
  965900101}

\bibitem[{Livi {et~al.}(2022)Livi, Larson, Kasper, Abiad, Case, Klein, Curtis,
  Dalton, Stevens, Korreck, {et~al.}}]{livi2022solar}
Livi, R., Larson, D.~E., Kasper, J.~C., {et~al.} 2022,
  \href{https://doi.org/10.3847/1538-4357/ac93f5}{APJ 938, 138.}

\bibitem[{Malaspina {et~al.}(2024)Malaspina, Ergun, Cairns, Short, Verniero,
  Cattell, \& Livi}]{malaspina2024frequency}
Malaspina, D.~M., Ergun, R.~E., Cairns, I.~H., {et~al.} 2024,
  \href{https://doi.org/10.3847/1538-4357/ad4b12}{APJ, 969, 60.}

\bibitem[{Mangeney {et~al.}(1999)Mangeney, Salem, Lacombe, Bougeret, Perche,
  Manning, Kellogg, Goetz, Monson, \& Bosqued}]{mangeney1999wind}
Mangeney, A., Salem, C., Lacombe, C., {et~al.} 1999,
  \href{https://doi.org/10.1007/s00585-999-0307-y}{Ann. Geophys., 17, 307-320}

\bibitem[{Marsch(1991)}]{Marsch1991}
Marsch, E. 1991, in Physics of the inner heliosphere II: Particles, waves and
  turbulence (\href{https://doi.org/10.1007/978-3-642-75364-0_3}{Springer}),
  \href{https://doi.org/10.1007/978--3--642--75364--0_3}{45--133}

\bibitem[{Montag {et~al.}(2025)Montag, Howes, McGinnis, Afshari, Starkey, \&
  Desai}]{montag2025mms}
Montag, P., Howes, G., McGinnis, D., {et~al.} 2025, The Astrophysical Journal
  Letters, 980, L23

\bibitem[{Montag \& Howes(2022)}]{montag2022field}
Montag, P., \& Howes, G.~G. 2022, Physics of plasmas, 29

\bibitem[{Mozer {et~al.}(2023{\natexlab{a}})Mozer, Bale, Kellogg, Romeo, Vasko,
  \& Verniero}]{Mozer2023a}
Mozer, F., Bale, S., Kellogg, P., {et~al.} 2023{\natexlab{a}},
  \href{https://doi.org/10.1063/5.0151423}{Phys. Plasmas, 062111, 30}

\bibitem[{Mozer {et~al.}(2023{\natexlab{b}})Mozer, Agapitov, Kasper, Livi,
  Romeo, \& Vasko}]{Mozer2023}
Mozer, F.~S., Agapitov, O.~V., Kasper, J.~C., {et~al.} 2023{\natexlab{b}},
  \href{https://doi.org/10.1051/0004-6361/202346202}{A\&A 673, L3.}

\bibitem[{Mozer {et~al.}(2022)Mozer, Bale, Cattell, Halekas, Vasko, Verniero,
  \& Kellogg}]{Mozer2022}
Mozer, F.~S., Bale, S.~D., Cattell, C.~A., {et~al.} 2022,
  \href{https://doi.org/10.3847/2041-8213/ac5520}{AJL 927, L15.}

\bibitem[{Mozer {et~al.}(2020)Mozer, Bonnell, Bowen, Schumm, \&
  Vasko}]{Mozer2020}
Mozer, F.~S., Bonnell, J.~W., Bowen, T.~A., Schumm, G., \& Vasko, I.~Y. 2020,
  \href{https://doi.org/10.3847/1538-4357/abafb4}{ApJ, 901, 107}

\bibitem[{Mozer {et~al.}(2021b)Mozer, Bonnell, Hanson, Gasque, \&
  Vasko}]{Mozer2021bb}
Mozer, F.~S., Bonnell, J.~W., Hanson, E. L.~M., Gasque, L.~C., \& Vasko, I.~Y.
  2021b, \href{https://doi.org/10.3847/1538-4357/abed52}{ApJ, 911, 89.}

\bibitem[{Mozer {et~al.}(2021)Mozer, Vasko, \& Verniero}]{mozer2021triggered}
Mozer, F.~S., Vasko, I.~Y., \& Verniero, J.~L. 2021,
  \href{https://doi.org/10.3847/2041-8213/ac2259}{ApJL, 919, L2.}

\bibitem[{M{\"u}ller {et~al.}(2020)M{\"u}ller, Cyr, Zouganelis, Gilbert,
  Marsden, Nieves-Chinchilla, Antonucci, Auchere, Berghmans, Horbury,
  {et~al.}}]{muller2020solar}
M{\"u}ller, D., Cyr, O.~S., Zouganelis, I., {et~al.} 2020,
  \href{https://doi.org/10.1051/0004-6361/202038467}{A\&A, 642, A1.}

\bibitem[{Nair {et~al.}(2025)Nair, Halekas, Cattell, Johnson, Hanson,
  Whittlesey, Larson, Livi, Kasper, Stevens, {et~al.}}]{nair2025suprathermal}
Nair, R., Halekas, J.~S., Cattell, C., {et~al.} 2025,
  \href{https://doi.org/10.3847/1538-4357/adbf1f}{ APJ, 984, 14.}

\bibitem[{{Perrone} {et~al.}(2011){Perrone}, {Valentini}, \&
  {Veltri}}]{Perrone2011}
{Perrone}, D., {Valentini}, F., \& {Veltri}, P. 2011,
  \href{http://dx.doi.org/10.1088/0004-637X/741/1/43}{APJ, 741, 43}

\bibitem[{Phan {et~al.}(2022)Phan, Verniero, Larson, Lavraud, Drake, Øieroset,
  Eastwood, Bale, Livi, Halekas, Whittlesey, Rahmati, Stansby, \&
  Pulupa}]{Phan2022}
Phan, T.~D., Verniero, J.~L., Larson, D., {et~al.} 2022,
  \href{https://doi.org/10.1029/2021GL096986}{Geophysical Research Letters, 49,
  e2021GL096986.}

\bibitem[{Píša {et~al.}(2021)Píša, Soucek, Santolík, Hanzelka, Nicolaou,
  Maksimovic, Bale, Chust, Khotyaintsev, \& Krasnoselskikh}]{Pisa2021}
Píša, D., Soucek, J., Santolík, O., {et~al.} 2021,
  \href{https://doi.org/10.1051/0004-6361/202140928}{A\&A 656, A14.}

\bibitem[{Shu \& Osher(1988)}]{SHU1988439}
Shu, C.-W., \& Osher, S. 1988,
  \href{https://doi.org/10.1016/0021-9991(88)90177-5}{J. Comput. Phys., 77,
  439-471.}

\bibitem[{Stix(1992)}]{stix1992waves}
Stix, T. 1992, \href{https://link.springer.com/book/9780883188590}{Waves in
  Plasmas (New York: American Institute of Physics)}

\bibitem[{Tenerani {et~al.}(2021)Tenerani, Sioulas, Matteini, Panasenco, Shi,
  \& Velli}]{tenerani2021evolution}
Tenerani, A., Sioulas, N., Matteini, L., {et~al.} 2021,
  \href{https://doi.org/10.3847/2041-8213/ac2606}{APJL, 919, L31.}

\bibitem[{Valentini {et~al.}(2011)Valentini, Perrone, \&
  Veltri}]{Valentini2011}
Valentini, F., Perrone, D., \& Veltri, P. 2011,
  \href{https://iopscience.iop.org/article/10.1088/0004-637X/739/1/54}{APJ 739,
  54.}

\bibitem[{Valentini {et~al.}(2014)Valentini, Vecchio, Donato1, Carbone1,
  Briand, Bougeret, \& Veltri}]{Valentini2014}
Valentini, F., Vecchio, A., Donato1, S., {et~al.} 2014,
  \href{https://iopscience.iop.org/article/10.1088/2041-8205/788/1/L16}{AJL
  788, L16.}

\bibitem[{Verniero {et~al.}(2021)Verniero, Howes, Stewart, \&
  Klein}]{Verniero2021}
Verniero, J., Howes, G., Stewart, D., \& Klein, K. 2021,
  \href{https://doi.org/10.1029/2020JA028361}{JGRA, 126, e2020JA028361}

\bibitem[{Verniero {et~al.}(2022)Verniero, Chandran, Larson, Paulson, Alterman,
  Badman, \& Bale}]{Verniero2022}
Verniero, J.~L., Chandran, B. D.~G., Larson, D.~E., {et~al.} 2022,
  \href{https://doi.org/10.3847/1538-4357/ac36d5}{ApJ, 924, 112}

\bibitem[{Verniero {et~al.}(2020)Verniero, Larson, Bowen, Livi, Bonnell,
  Rahmati, Alterman, McManus, Whittlesey, Pyakurel, Malaspina, \&
  Klein}]{Verniero2020}
Verniero, J.~L., Larson, D., Bowen, T.~A., {et~al.} 2020,
  \href{https://doi.org/10.3847/1538-4365/ab86af}{ApJS, 248, 5}

\bibitem[{Wilson~III {et~al.}(2007)Wilson~III, Cattell, Kellogg, Goetz,
  Kersten, Hanson, \& MacGregor}]{Wilson2007}
Wilson~III, L., Cattell, C., Kellogg, P., {et~al.} 2007,
  \href{https://doi.org/10.1103/PhysRevLett.99.041101}{PhRvL, 99, 041101}

\end{thebibliography}

\end{document}